\documentclass[]{mn2e}

\usepackage{amssymb}
\usepackage{amsmath}
\usepackage{graphicx, subfigure}
\usepackage{appendix}
\usepackage{longtable}
\usepackage{url}
\title[PyMorph and SDSS photometry: sky and model fitting]{Comparing PyMorph and SDSS photometry. I.\\
  Background sky and model fitting effects}
\author[Fischer et al.]{\parbox{\textwidth}{J.-L. Fischer$^{1}$\thanks{E-mail: jofis@sas.upenn.edu}, M. Bernardi$^{1}$ \& A. Meert$^{1}$} \vspace{0.4cm}\\
\parbox{\textwidth}{$^{1}$Department of Physics and Astronomy, University of Pennsylvania, Philadelphia, PA 19104, USA\\}}

\begin{document}
 %\date{Accepted .  Received ; in original form }

\maketitle

\label{firstpage}

\begin{abstract}
  A number of recent estimates of the total luminosities of galaxies in the SDSS are significantly larger than those reported by the SDSS pipeline.  This is because of a combination of three effects:  one is simply a matter of defining the scale out to which one integrates the fit when defining the total luminosity, and amounts on average to $\le 0.1$~mags even for the most luminous galaxies.  The other two are less trivial and tend to be larger; they are due to differences in how the background sky is estimated and what model is fit to the surface brightness profile.
  We show that PyMorph sky estimates are fainter than those of the SDSS DR7 or DR9 pipelines, but are in excellent agreement with the estimates of Blanton et al. (2011).
  %Use of the SDSS sky biases luminosities, half-light radii, Sersic indices, and B/T ratios to lower values. 
  Using the SDSS sky biases luminosities by more than a few tenths of a magnitude for objects with half-light radii $\ge 7$ arcseconds.  In the SDSS main galaxy sample these are typically luminous galaxies, so they are not necessarily nearby.  This bias becomes worse when allowing the model more freedom to fit the surface brightness profile.     When PyMorph sky values are used, then two component Sersic-Exponential fits to E+S0s return more light than single component deVaucouleurs fits (up to $\sim 0.2$~mag), but less light than single Sersic fits ($0.1$~mag).  
  Finally, we show that PyMorph fits of Meert et al. (2015) to DR7 data remain valid for DR9 images.  Our findings show that, especially at large luminosities, these PyMorph estimates should be preferred to the SDSS pipeline values.
%and that estimates based on individual objects are reliable.
%; stacking is not a prerequisite for obtaining unbiased results.
   
\end{abstract}

\begin{keywords}
 galaxies: fundamental parameters -- galaxies: photometry -- galaxies: structure
\end{keywords}

\section{Introduction}
There is substantial interest in quantifying the luminosity and stellar mass functions in the local universe (Bernardi et al. 2017a and references therein).  The Sloan Digital Sky Survey (hereafter SDSS), which surveyed about a quarter of the sky to a median redshift of about $z\sim 0.1$, is the benchmark database for such studies.  Recently Meert et al. (2015, 2016) have made available a re-analysis of the galaxies in the SDSS DR7 release (Abazajian et al. 2009).  Their analysis determines photometric parameters, such as luminosity, half-light radius, a measure of the steepness or central concentration of the profile, etc., by fitting a number of different models to the surface brightness profile:  a single component deVaucouleurs profile, a single component Sersic profile, and a two component Sersic bulge plus exponential disk profile (hereafter deV, Ser and SerExp).  The fitting algorithm is called PyMorph (Vikram et al. 2010; Meert et al. 2013, 2015, 2016; Bernardi et al. 2014).  The PyMorph catalog yields substantially more light at high luminosities (Bernardi et al. 2013, 2016a,b, and Figure~\ref{pymorphSDSS7} below) than previous work based on SDSS pipeline photometry.  The differences impact Halo Model (Cooray \& Sheth 2002) based interpretations of the relationship between galaxies and dark matter halos at $z\sim 0.1$ (e.g. Shankar et al. 2014).  Pinning down this relationship locally is crucial for studies of how this relationship evolves. 

In addition, as first identified by Bernardi et al. (2011) at the high mass (luminosity) end there is a special mass (luminosity) scale: $2\times 10^{11}M_\odot$ (which corresponds to an $r-$band luminosity scale of $\sim -22.5$ mag). Various scaling relations change slope at this scale, and this is thought to be related to a change in the assembly histories -- e.g. minor versus major dry mergers. It is also the mass (or luminosity) scale where the stellar mass (or luminosity) function starts to drop exponentially. For all these reasons, identifying and accounting for all possible biases so as to have reliable photometric estimates at these luminosity and mass scales is important.  Here we address the reasons for the differences between PyMorph and the SDSS, and show that PyMorph should be used, especially at large luminosities.

\begin{figure}
 \centering
         \includegraphics[width = 0.9\hsize]{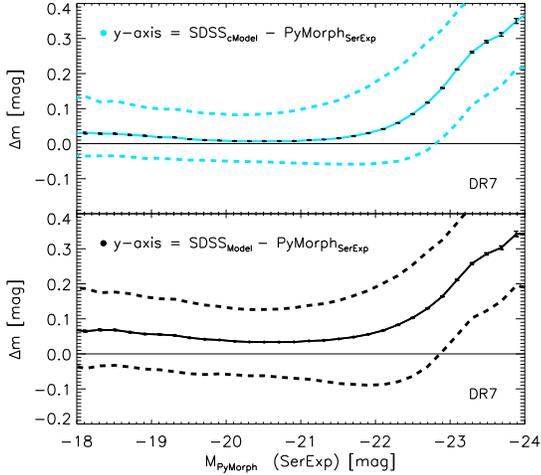}
 \caption{Difference in the total light estimated by the SDSS and PyMorph (i.e. Meert et al. 2015) in DR7.  Solid line with error bars shows the median in each bin in magnitude and dashed lines show the region which contains 68\% of the objects in the bin.  Top panel shows that SDSS DR7 {\tt cModel} magnitudes and PyMorph DR7 SerExp magnitudes are similar except for the most luminous galaxies, where PyMorph is brighter.  Bottom panel shows the result of replacing {\tt cModel} with {\tt Model} magnitudes; except for an overall offset, the trends are similar.}
 \label{pymorphSDSS7}
\end{figure}

There are expected to be three main culprits.  An important step in the determination of the amount of light we receive from an object is the estimation of the amount of light which is contributed by the background sky.  Over-estimating the contribution from the sky will lead to an underestimate of the size and total light, and perhaps a decrease in the estimate of how centrally concentrated the object is.  Bernardi et al. (2007) (see also, e.g., SDSS DR7 documentation) noted that the SDSS pipeline reductions underestimated the sky, especially in crowded fields.  In the years since, the SDSS has revised its pipelines (see the DR9, Ahn et al. 2012, and subsequent data releases).  

In addition, a number of other analyses have also provided improved estimates (Simard et al. 2011, Blanton et al. 2011, Meert et al. 2015, 2016).  One of the main goals of the present work is to compare different estimates of the sky in the SDSS footprint, and to quantify the impact this has on the estimated sizes, shapes and luminosities of galaxies.  Blanton et al. (2011) argue that the SDSS values can be biased by as much as a magnitude for nearby objects with large angular size (half-light radius $\ge 40$ arcseconds).  However, because the bias is really associated with having a large angular size, the bias can still be significant (a few tenths of a magnitude) for large objects (half-light radius $\ge 7$ arcseconds) whether or not they are nearby.  There is a tight correlation between luminosity and physical size, so even though the majority of luminous galaxies in the SDSS main galaxy sample tend to be more distant ($z \sim 0.2$) they still have relatively large angular sizes ($\ge 7$ arcseconds).

\begin{figure}
 \centering
	 \includegraphics[width = 0.9\hsize]{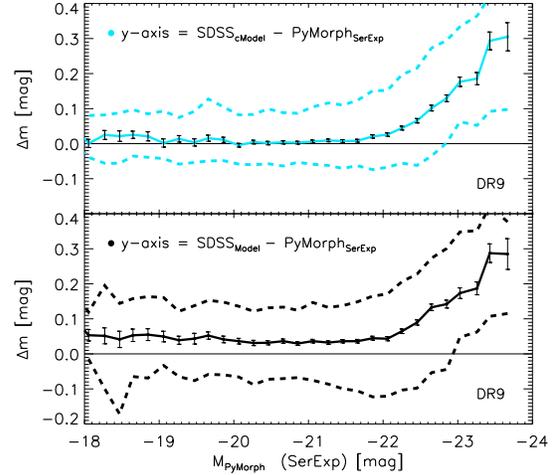}
 \caption{Same as Figure~\ref{pymorphSDSS7}, but now using DR9 values for the subset of $10^4$ objects which were used in Meert et al. (2013) to test PyMorph and for which we re-ran PyMorph using the DR9 images.}
 \label{pymorphSDSS9}
\end{figure}

In addition to the sky, two other effects contribute to differences between SDSS pipeline and more recent estimates of galaxy luminosities and sizes.  One is trivial:  when reporting the total light in an image, the SDSS only integrates the surface brightness profile out to about $\sim 7\times$ the half-light radius.  Others, such as PyMorph (Meert et al. 2013), do not truncate.  This amounts to a small systematic difference of order 0.05~mags for deV profiles, but can be larger for Ser profiles (e.g. Kelvin et al. 2012).  The second effect is more interesting:  it is the fact that the luminosity and size estimates depend on the model which is fitted to the image.  In what follows, we will be careful to distinguish between these three effects.  E.g., it is not obvious if models which have more freedom to better fit the image will end up predicting more light or less.

There is another potential observational systematic:  the deblending of overlapping galaxies.  However, this is resolved in Meert et al. (2015), who discuss how PyMorph handles nearby neighbours, as well as polluted fits (those that could not be deblended).  Their rate of occurence is sub-percent, and PyMorph provides a flag identifying them, so it was simple to exclude them from the analysis which follows.  

The present study is timely because the Meert et al. analyses are based on SDSS DR7 images.  However, significant changes to the SDSS imaging pipeline were implemented in DR9, and remain in place in subsequent data releases.  These are described on the SDSS website: {\tt www.sdss.org}.  Therefore, after defining the sample we work with in Section~\ref{sample}, our first step is to compare PyMorph analyses of the DR7 and DR9 images.  This is the subject of Section~\ref{sdss2->3}.  Section~\ref{trunc} and Section~\ref{bimodal} quantify the effects of truncation.  Section~\ref{bimodal} also highlights the fact that, because the most massive objects may be a different population having different profile shapes it is important to specify the choice of regression, i.e. the average magnitude difference may depend on the luminosity being used as x-axis.  Section~\ref{sky} compares sky estimates from the SDSS DR7 and DR9 pipelines with determinations from Blanton et al. (2011), Simard et al. (2011), and Meert et al. (2015, 2016) (hereafter B11, Simard11 and PyMorph DR7, respectively).  Section~\ref{Pmodels} shows how the choice of model to fit affects the estimated total light.  A final section summarizes.

When necessary, we assume a spatially flat background cosmology with parameters $(\Omega_m,\Omega_\Lambda)=(0.3,0.7)$, and a Hubble constant at the present time of $H_0=70$~km~s$^{-1}$Mpc$^{-1}$.

\begin{figure}
 \centering
	\includegraphics[width = 0.9\hsize]{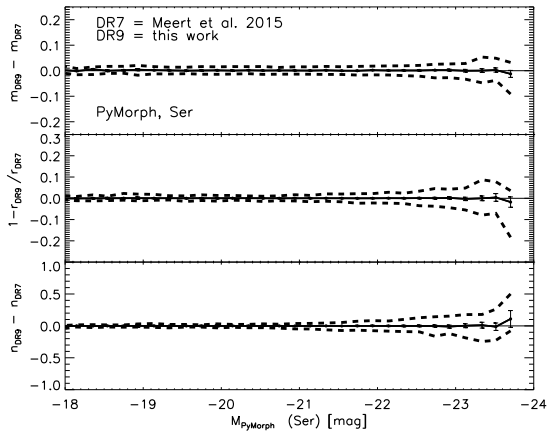}
 \vspace{-1cm}
 \caption{Comparison of PyMorph Sersic photometric parameters in DR7 (Meert et al. 2015) and DR9 (this work), showing that apparent magnitudes, half-light radii, and Sersic indices are essentially unchanged.  PyMorph DR7 parameters are also valid for DR9.}
 \label{dr7to9ser}
\end{figure}

\section{Comparison of SDSS and PyMorph}\label{sdssPymorph}
The analysis which follows is based on the SDSS DR7 and DR9 Main Galaxy samples.  For these galaxies, the SDSS provides a number of photometric parameters on its website:  {\tt www.sdss.org}.  We are most interested in the total magnitudes and half-light radii, the best SDSS pipeline estimates of which are based on fitting exponential or deVaucouleurs profiles to the sky subtracted image.  {\tt Model} magnitudes simply choose the better of the two fits, whereas {\tt cModel} magnitudes use a linear combination of the two best fits (a $\chi^2$-like goodness of fit metric is minimized to set the relative amplitudes of the components).  Thus, although they are the result of fitting two profile shapes, {\tt cModel} magnitudes are not really two-component fits.  

In contrast to the SDSS {\tt cModel} photometry, the best PyMorph SerExp photometry is based on true two-component fits -- a Sersic bulge with an exponential second component -- in which the sky, assumed to be constant across the image, is also fit simultaneously (e.g. Meert et al. 2015).  These fits were made using the DR7 release.  

\subsection{Motivation}\label{sample}
We begin with a comparison of what are considered to be the best SDSS and PyMorph photometry:  {\tt cModel} and SerExp magnitudes.  
%The former are a linear combination of the best fitting exponential and the best fitting deVaucouleur profile to the surface brightness profile of a sky-subtracted image; the latter are a true two-component fit (a Sersic bulge with an exponential second component) in which the sky (assumed to be constant across the image) is also fit simultaneously.  
Figure~\ref{pymorphSDSS7} shows that the two are in good agreement, except at the bright end, where PyMorph is substantially brighter.  The bottom panel shows the result of replacing {\tt cModel} with {\tt Model} magnitudes.
%; these simply choose the better fitting of the exponential and deVaucouleur profiles without forming a linear combination of the two fits.  
Except for an offset at low and intermediate luminosities, both panels show similar trends.  The similarity observed at the bright end is expected because the vast majority of the most luminous galaxies are E+S0s, so {\tt Model} = {\tt deV} and {\tt cModel} $\approx$ {\tt deV}.
% (e.g. Figure~\ref{deVcModMod}). 
 
The main goal of the present study is to determine which of the three culprits mentioned in the Introduction are responsible for the offsets in Figure~\ref{pymorphSDSS7}.  In particular, it may be that the agreement between {\tt cModel} and SerExp at faint and intermediate luminosities is fortuitous.

\begin{figure}
 \centering
 	\includegraphics[width = 0.9\hsize]{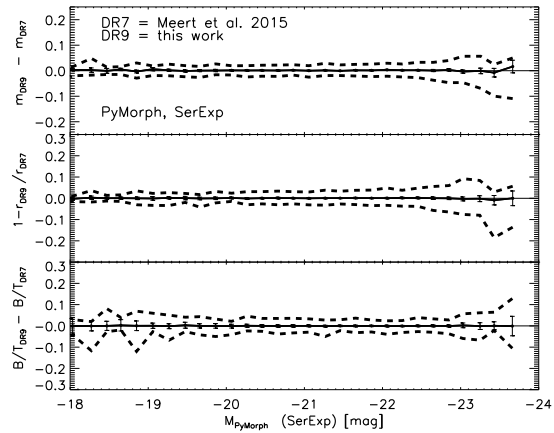}
 \vspace{-1cm}
 \caption{Same as previous figure, but now for PyMorph SerExp fits.  Since there are now two components, the panel showing Sersic index has been replaced by one showing B/T ratio.  Again, PyMorph DR7 parameters are valid for DR9, although the scatter around the median is larger compared to Figure~\ref{dr7to9ser}, since with more free parameters, there can be more degeneracies between the best fit values.  
}
 \label{dr7to9serexp}
\end{figure}

Figure~\ref{pymorphSDSS7} was made using DR7 galaxies.  However, between DR7 and DR9, a number of parts of the SDSS pipeline were changed.  The most important change is the SDSS sky estimate, but how flux calibration is done, and so on, also changed (see Aihara et al. 2011 and Ahn et al. 2012 for details).  Therefore, our first step is to determine if the changes from DR7 to DR9 matter.  Figure~\ref{pymorphSDSS9} shows a similar comparison as in Figure~\ref{pymorphSDSS7}, but now using DR9 values.  To make this figure we ran PyMorph on a subset of $10^4$ DR9 galaxies.  The chosen objects are the same as those used by Meert et al. (2013) when developing and testing PyMorph. The distribution of the measured parameters of this subset reproduces the distribution of all the observed galaxies in the SDSS DR7 main galaxy sample (see their Figure~1). Comparison of Figure~\ref{pymorphSDSS9} with Figure~\ref{pymorphSDSS7} shows little difference:  the discrepancy between SDSS and PyMorph which was known to exist in DR7 persists in DR9.  

\subsection{Comparison of SDSS DR7 and DR9}\label{sdss2->3}
We now consider if the best fitting PyMorph parameters have changed between DR7 and DR9.  Since PyMorph fits for the sky itself -- it does not use the SDSS value -- we expect the change to the SDSS sky estimate to have little impact on the PyMorph fits.  Figure~\ref{dr7to9ser} shows that this is indeed the case:  the apparent magnitudes, sizes, and Sersic indices for PyMorph Ser fits are essentially unchanged.  Figure~\ref{dr7to9serexp} shows that this is also true for PyMorph SerExp fits; because these are two-component fits, the bottom panel shows bulge/total ratios rather than Sersic indices. Both figures show that, although there is scatter between the DR7 and DR9 values, it is similar to the statistical uncertainty on the parameters (Meert et al. 2013). It should be noted that there is larger scatter for SerExp than for Ser, because there are more free parameters and hence more potential degeneracies. Therefore, Figures~\ref{dr7to9ser} and~\ref{dr7to9serexp} indicate that the PyMorph parameters of Meert et al. can be used essentially without modification even for DR9.  (There are, of course, other studies for which the difference between DR7 and DR9 or DR13 recalibrations do matter.)

\begin{figure}
 \centering
	\includegraphics[width = 0.9\hsize]{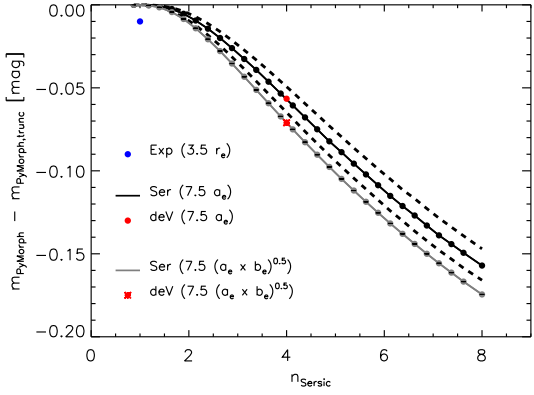}
 \caption{Effect of truncation on the value of the reported total magnitude, as a function of Sersic index $n$.  When the truncation radius is a fixed multiple of $\sqrt{ab}$, then this difference does not depend on $a$ or $b$, and the effect of truncation is the same as truncating a spherical profile with $r_e=\sqrt{ab}$ (grey solid line).  However, if it is a multiple of $a$, then this difference depends on $b/a$.  Black solid curve shows the median value of this difference for SDSS E+S0 galaxies and dashed curves show the region which enclose 68\% of the values at each $n$.}
 \label{ftruncate}
\end{figure}

\subsection{Effect of truncation}\label{trunc}
In what follows, we would like to compare the luminosity estimates of PyMorph and the SDSS.  Both report values based on fitted models; however, whereas PyMorph integrates the fitted profile to infinity, the SDSS does not.  If a two-dimensional Sersic profile with semi-major axis $a$ and axis ratio $b/a$ is truncated along a line of constant surface brightness, then 
\begin{equation}
 L_{\rm trunc} = L_\infty\,\frac{\gamma(2n, b_n\,\rho_{\rm trunc}^{1/n})}{\Gamma(2n)},
 \ \ {\rm where}\ \ 
 \rho_{\rm trunc} \equiv \frac{\theta_{\rm trunc}}{\sqrt{ab}},
 \label{Ltrunc}
\end{equation} 
$\gamma(m,x)$ is the incomplete gamma function, $\gamma(m,\infty) = \Gamma(m)$, and $b_n$ is defined by requiring $\gamma(2n,b_n)=\Gamma(2n)/2$.  E.g., $b_n\approx 7.669$ when $n=4$.

\begin{figure}
 \centering
	\includegraphics[width = 0.9\hsize]{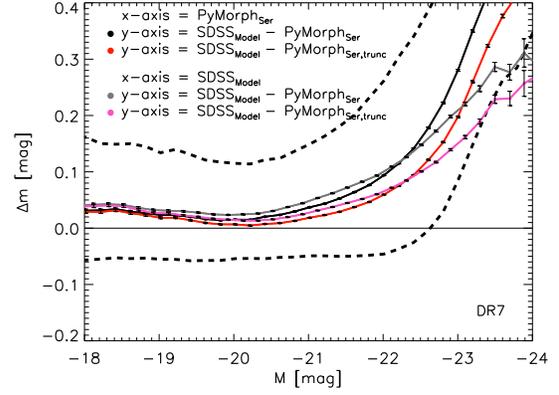}
 \caption{Magnitude differences shown as a function of {\tt Ser}\ and SDSS {\tt Model}\ magnitudes.  At the bright end, the choice matters; while using {\tt Ser}\ magnitudes truncated at $7.5a$ reduces the difference (truncating at $7.5\sqrt{ab}$ is almost identical to $7.5a$), it is still true that plotting versus {\tt Ser}\ instead of SDSS {\tt Model}\ magnitudes returns a substantially larger value for the mean difference at the bright end.  }
 \label{curved1}
\end{figure}

\begin{figure}
 \centering
	\includegraphics[width = 0.9\hsize]{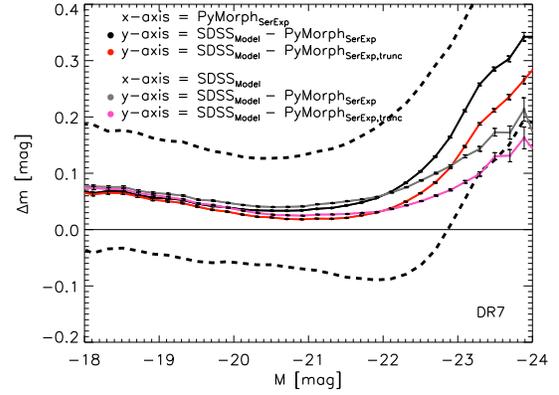}
 \caption{Same as previous figure, but now for {\tt SerExp}\ rather than {\tt Ser}\ photometry.}
 \label{curved2}
\end{figure}

The ratio $L_{\rm trunc}/L_\infty$ clearly depends on $n$.  Notice that if $\theta_{\rm trunc}$ is a multiple of $\sqrt{ab}$, then, at fixed $n$, the correction is the same for all axis-ratios. For example, in their work with the GAMA survey, Kelvin et al. (2012) set $\theta_{\rm trunc} = 10\sqrt{ab}$. For reasons which will become clear shortly, the grey solid curve in Figure~\ref{ftruncate} shows $\theta_{\rm trunc}=7.5\sqrt{ab}$. This shows that, when $n=4$, the correction is 0.07~mags. Unfortunately, the SDSS truncation is more complicated:  the SDSS website says that it truncates with a function which drops from unity to zero between $7$ and $8\times$ the half-light radius.  However, in the database, the quantity which is called $r_e$ is the semi-major axis $a$, rather than $\sqrt{ab}$.  In addition, the actual form of this truncation has never been published.  As we show below, we are able to reproduce the SDSS values if we use a sharp truncation radius of $7.5a$ making $\rho_{\rm trunc}^{\rm SDSS} \approx 7.5a/\sqrt{ab} = 7.5\,\sqrt{a/b}$.  (In particular, $7.5a$ works substantially better than $7.5\sqrt{ab}$.)  Hence, at fixed $n$, $L_{\rm trunc}/L_\infty$ is a monotonic function of $b/a$:  since $0\le b/a \le 1$, the correction is maximal when $b/a=1$ and $L_{\rm trunc}\to L_\infty$ as $b/a\to 0$.  Thus, at fixed $n$, there is a range of corrections which depends on the distribution of $b/a$.  Since our goal is to compare with the SDSS, the black solid line in Figure~\ref{ftruncate} shows the median of $2.5\log_{10}(L_{\rm trunc}/L_\infty)$ as a function $n$, and the scatter around this median (black dashed lines), for the PyMorph Sersic reductions of SDSS E+S0 galaxies when $\theta_{\rm trunc}=7.5a$.  This shows that, when $n=4$, the correction is $\sim 0.05$~mags, but when $n=8$, then the median correction is $\sim 0.16$~mags. (For later type galaxies $n$ is smaller so the correction is smaller; the blue dot shows the correction if $n=1$ and one truncates at $3.5\times$ the half light radius.) 

In what follows, we will be careful to indicate if the reported magnitudes were based on truncation or not.  However, the half-light radii we report are always those which include $L_\infty/2$; we never use the scale associated with $L_{\rm trunc}/2$.

\subsection{Choice of regression and truncation}\label{bimodal}

We remarked in the Introduction that the most massive galaxies appear to be a structurally different population. So it should not be surprising if their surface brightness profiles are also different in some way.  If these are objects for which SDSS and PyMorph photometry is particularly different, then plots versus {\tt PyMorph}\ may look rather different from plots versus {\tt Model}, for the same reason that, in a Gaussian mixture model, plots of $y$ versus $x$ can look very different from plots of $x$ vs $y$.
%Therefore, if one wishes to use the mean of $\Delta M$ as a measure of the difference between {\tt Model}\ and {\tt Ser}, then one {\em must} specify which variable was held fixed.  
Figure~\ref{curved1} shows that something like this happens in the SDSS data:  The differences between {\tt Ser}\ and {\tt Model}\ magnitudes increase at the bright end, but they look much larger when shown as a function of {\tt Ser}\ rather than {\tt Model}\ magnitudes. Figure~\ref{curved2} shows that the same is true of {\tt SerExp}\ magnitudes.

  The differences are reduced slightly if one uses truncated {\tt Ser}\ or {\tt SerExp}\ magnitudes, since this reduces the analog of $m$ in the example above, but it does not change the fact that the choice of $x$-axis matters.  While truncation matters, the net effect of truncation is about half of what one would naively have expected from Figure~\ref{ftruncate}. This is because the correction depends on $n$, but the $n$-$L$ correlation is weak. Although large $L$ have larger $n$, so truncation matters more for large $L$, there is substantial scatter around the mean $n$ which reduces the net effect.  This is also why, in practice, it matters little ($\le 0.01$~mags) whether one truncates using $7.5a$ or $7.5\sqrt{ab}$. Of course, truncation matters even less for the SerExp fits.

\section{Comparison of sky estimates}\label{sky}
\begin{figure}
 \centering
	\includegraphics[width = 0.9\hsize]{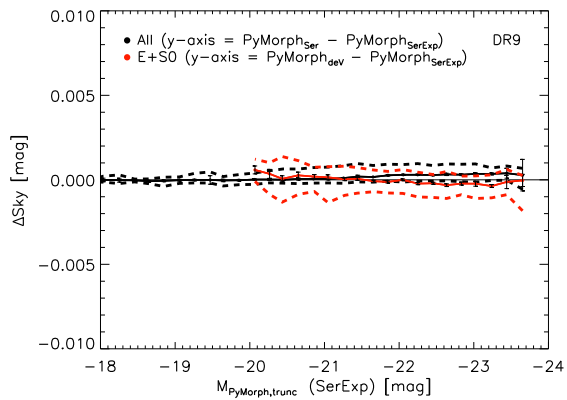}
 \caption{Comparison of PyMorph DR9 sky estimates from Sersic and SerExp fits to images of all galaxies, and from deVaucouleur and SerExp fits to E+S0s.  The estimated sky depends very weakly on which model is fitted to the image.  Therefore, in what follows, we use the PyMorph SerExp sky as representative of all PyMorph sky values.}
 \label{sameSky}
\end{figure}

Our goal is to compare PyMorph and SDSS sky estimates.  However, when fitting a model to the observed galaxy image, PyMorph fits for the sky -- assumed to have constant surface brightness across the image -- simultaneously.  Therefore, it is possible that the fitted sky varies when the model which is fitted to the galaxy surface brightness profile changes.  This would make comparisons with the SDSS sky estimate depend on the fitted model.  Fortunately, Figure~\ref{sameSky} shows that the estimated sky is essentially the same whatever the fitted model.  (We have plotted versus truncated magnitudes. Of course, the y-axis legend does not specify truncated because the sky estimates do not depend on (i.e. are the same) whether or not we truncate.)

This has two consequences.  First, when comparing the PyMorph sky with other estimates, we do not need to specify if it is the deV sky, the Ser sky, or the SerExp sky, since, for the present purposes, they are all the same.  We exploit this fact in Section~\ref{skydeV}.  Second, the similarity in sky values indicates that differences between PyMorph models are not driven by the sky.  We use this fact in Section~\ref{Pmodels}.  

\begin{figure*}
 \centering
	\includegraphics[width = 0.45\hsize]{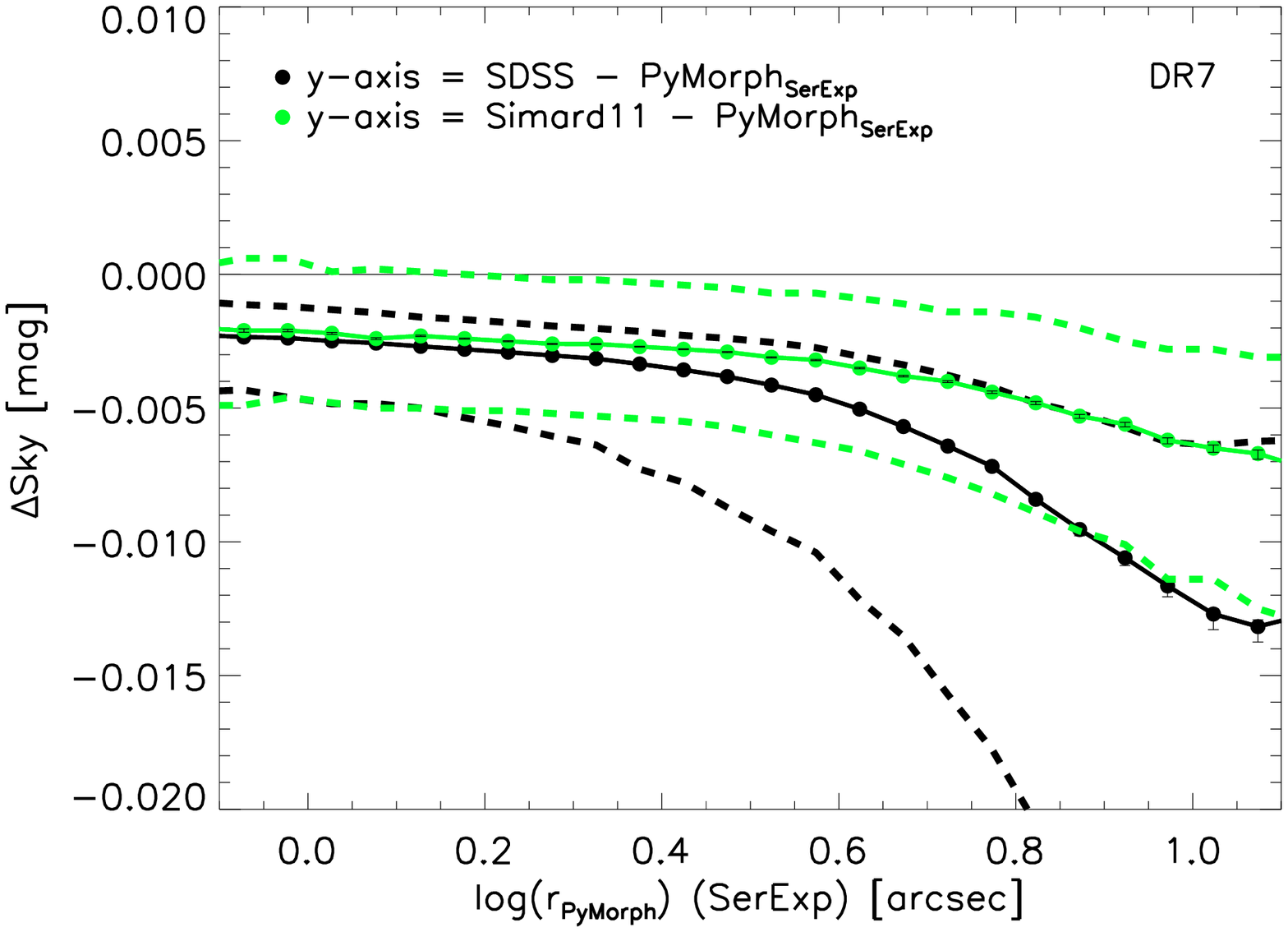}
	\includegraphics[width = 0.45\hsize]{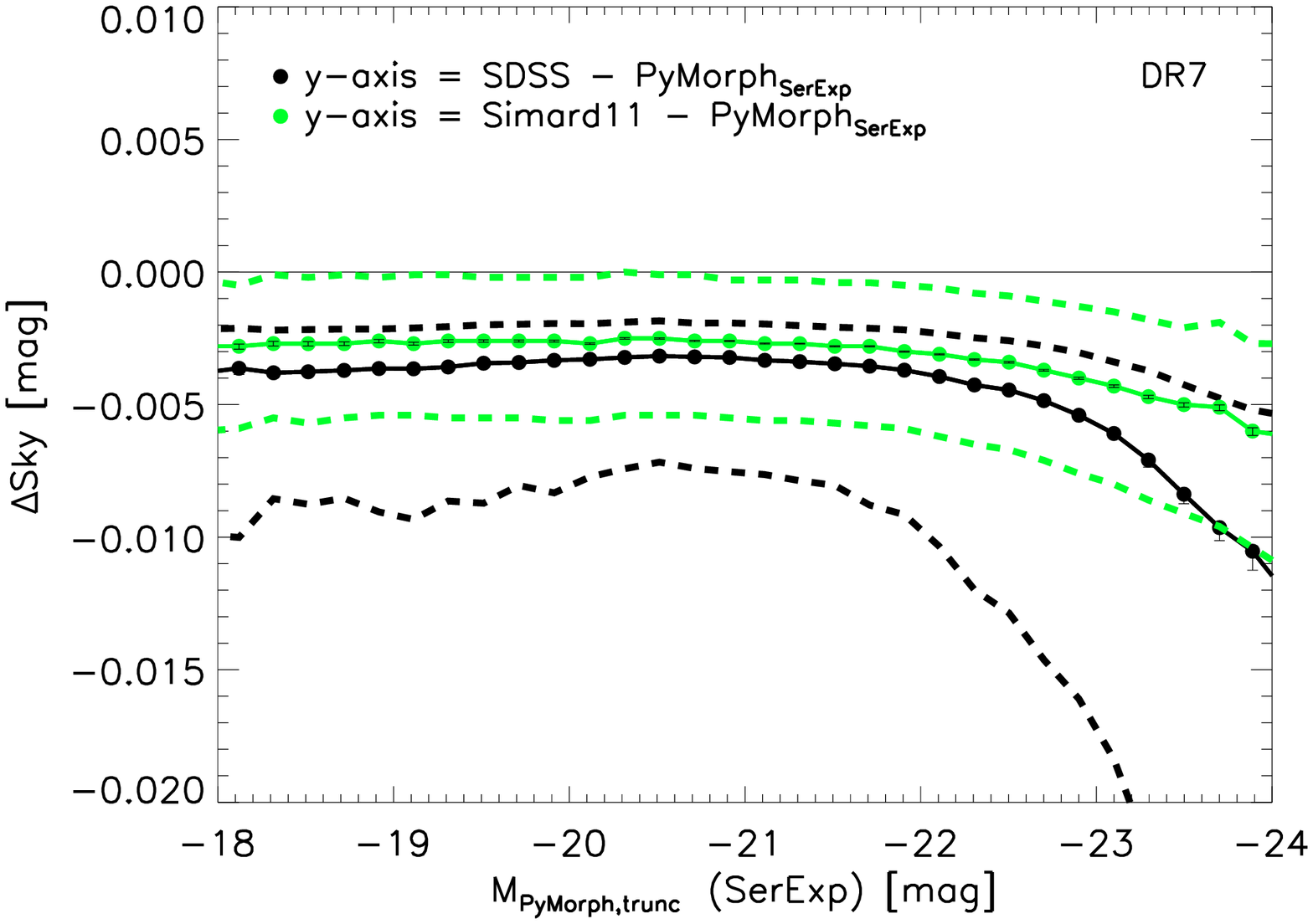}
 \caption{Comparison of background sky estimates from SDSS DR7, Simard11, and PyMorph DR7 SerExp (i.e. Meert et al. 2015), shown as a function of apparent size (left) and absolute magnitude (right).  The PyMorph sky is faintest and SDSS brightest, particularly for objects with large angular sizes and/or luminosities.  The Simard11 sky lies closer to the SDSS than to PyMorph.}
 \label{SDSS2sky}
\end{figure*}

\begin{figure*}
 \centering
	\includegraphics[width = 0.45\hsize]{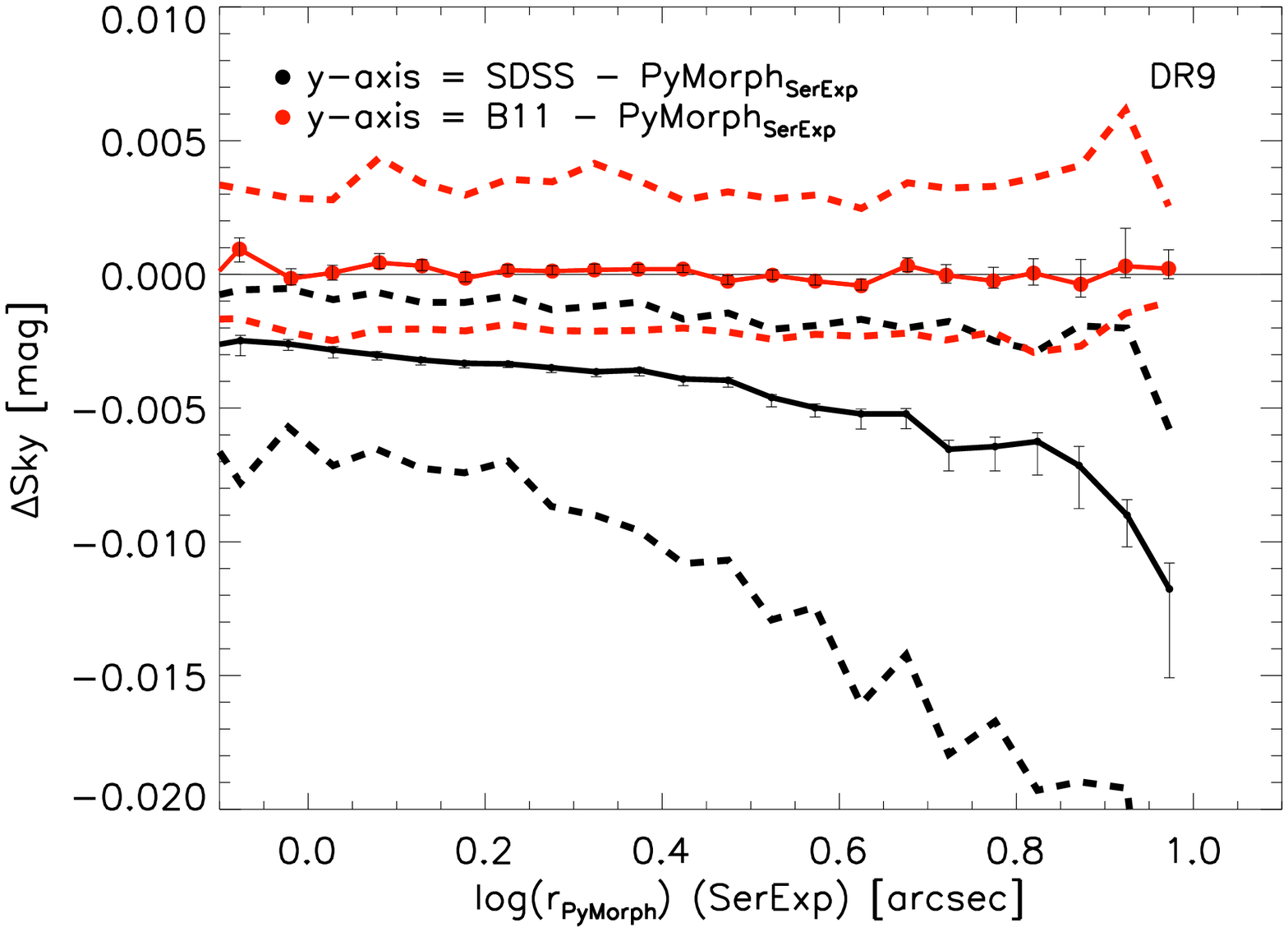}
	\includegraphics[width = 0.45\hsize]{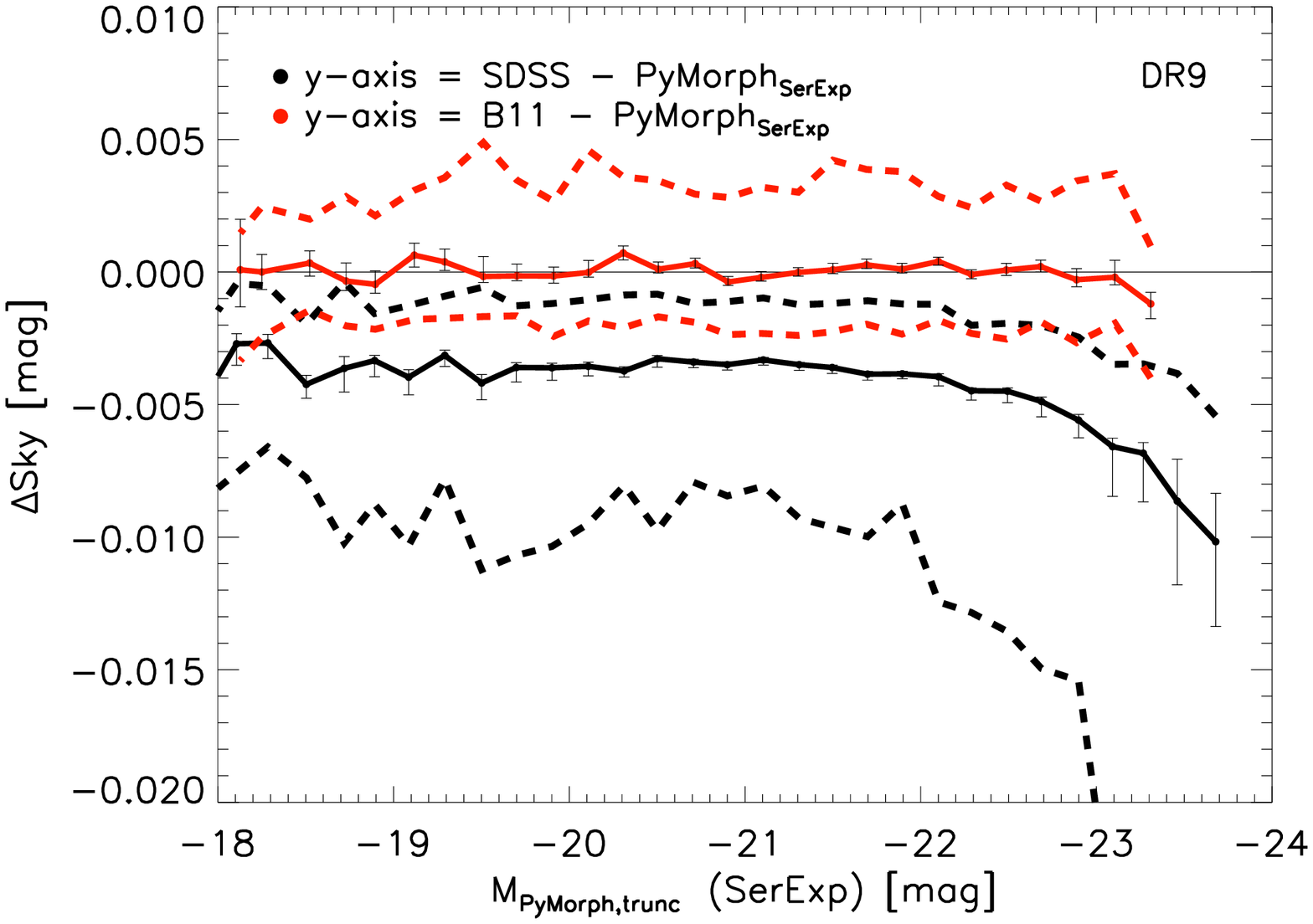}
 \caption{Comparison of background sky estimates from SDSS DR9, Blanton et al. (2011; B11), and PyMorph DR9 SerExp (i.e. this work), for the same subset of $10^4$ galaxies which were used to make Figure~\ref{pymorphSDSS9}.  Left and right panels show results as a function of angular size and luminosity.  B11 and PyMorph are in excellent agreement with one another, whereas SDSS DR9 is clearly brighter.}
 \label{blantonSky}
\end{figure*}

\begin{figure*}
 \centering
 	 \includegraphics[width = 0.45\hsize]{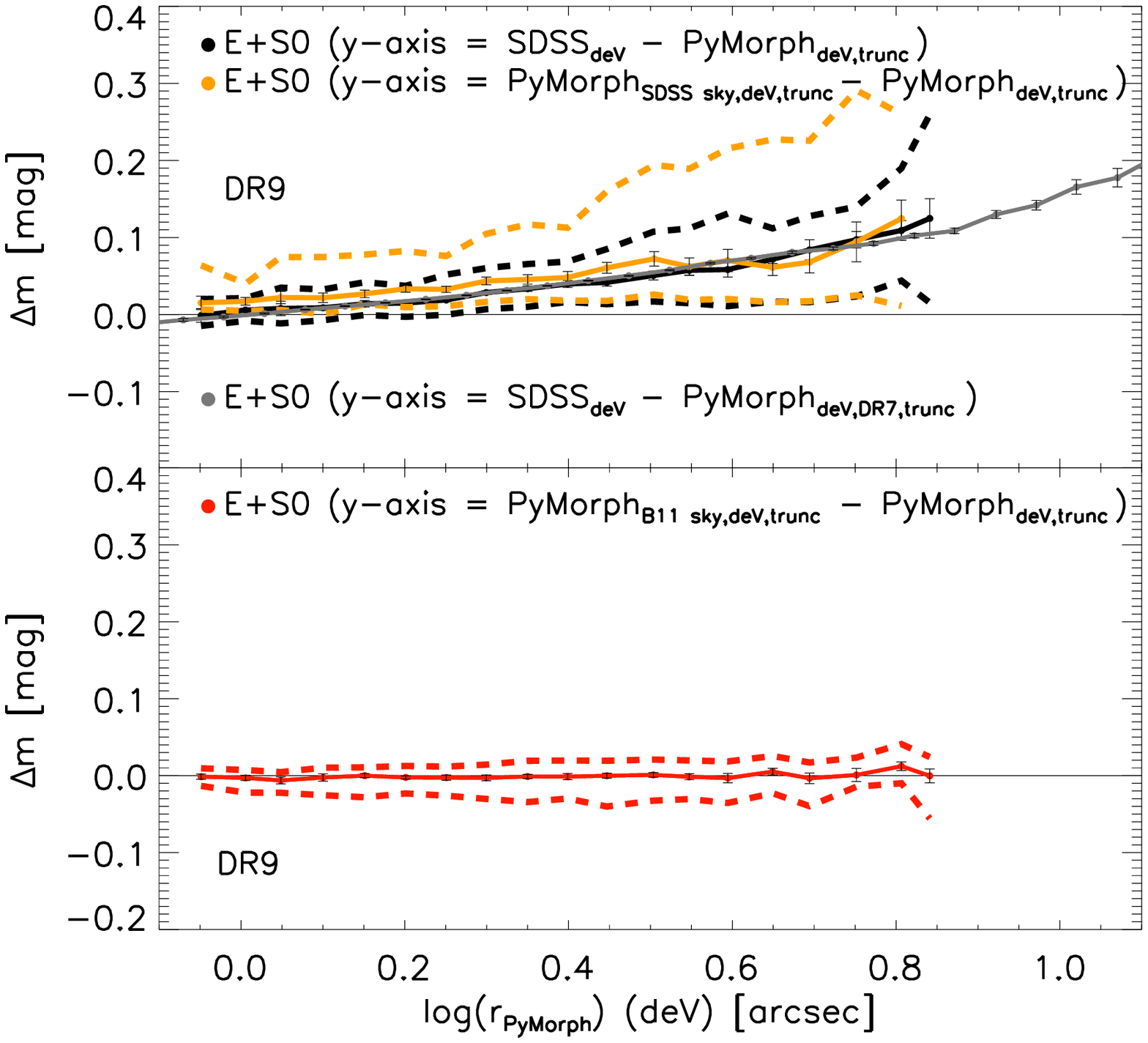}
	 \includegraphics[width = 0.45\hsize]{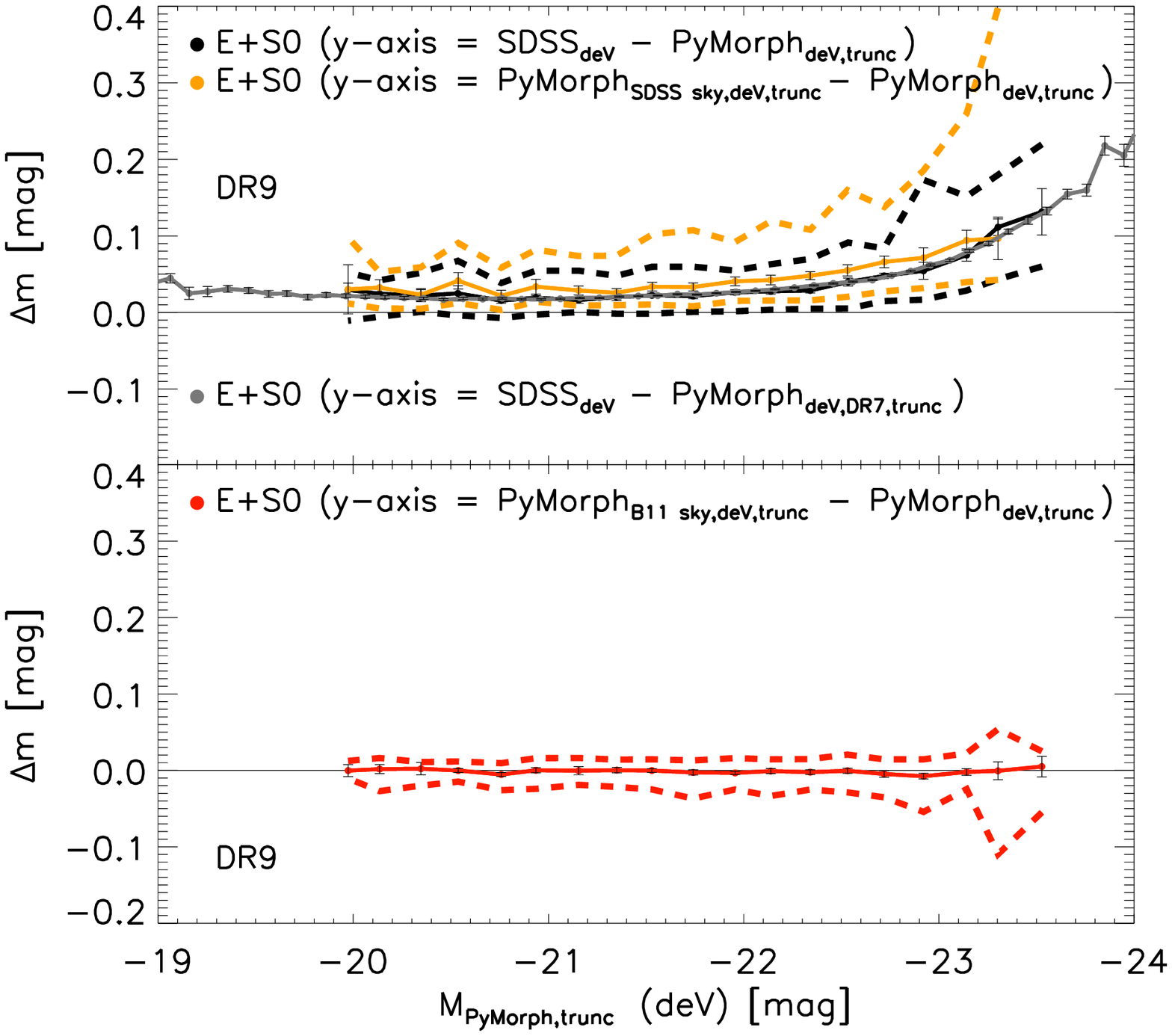}
 \caption{Difference between the SDSS DR9 and PyMorph DR9 estimates of the total (truncated) light, based on fitting a deVaucouleur profile to the image of an E or S0 galaxy.  Black symbols and curves show results when PyMorph is allowed to fit its own sky on the DR9 subset of galaxies; yellow symbols and curves are when it is forced to use the SDSS sky value. The agreement between the black and yellow curves suggests that {\tt PyMorph}$_{\tt SDSSsky,deV,trunc}$ is a good proxy for {\tt SDSS}$_{\tt deV}$.
The gray curve shows the comparison of the much larger ($\sim 60\times$) sample between SDSS DR9 and PyMorph DR7 (i.e. Meert et al. 2015). The SDSS estimate is systematically fainter and this difference increases for the galaxies with the largest angular sizes (left) or luminosities (right).  Bottom set of panels shows a similar comparison of PyMorph estimates when the sky is fixed to that of Blanton et al. (2011; B11) and when it is not:  the differences are negligible and the scatter is smaller compared to the top panels. }
 \label{deV}
\end{figure*}

We are now ready to compare background sky estimates with those from PyMorph, for which we use the SerExp sky value.  Figure~\ref{SDSS2sky} compares background sky estimates from SDSS DR7, Simard11, and PyMorph DR7 SerExp.  The PyMorph sky is faintest and SDSS brightest, with the Simard11 sky lying closer to the SDSS at the faint and intermediate luminosities and in between at the bright end.  The differences from PyMorph are particularly large for objects with large angular sizes or luminosities.

While it is tempting to conclude that Simard11 is the most prudent choice because it lies between the other two, Figure~\ref{blantonSky} shows that the PyMorph sky estimate is in excellent agreement with that of B11. In contrast to the previous figure, this one uses DR9 images, for which Simard11 values are not available.  

The PyMorph and B11 sky values were determined in very different ways.  Those of B11 are based on fitting the masked background sky for each SDSS scan with a smooth continuous function across the sky.  (In Figure~\ref{blantonSky}, we used the B11 sky value measured at the center of the galaxy image since the variation of the sky value on the scale of a galaxy is very small.)  In contrast, the PyMorph sky is determined on an object-by-object basis.  
%For instance, whereas PyMorph assumes the sky is constant across a galaxy, the B11 estimate can vary.  
Therefore, the agreement between the two is nontrivial, and strongly suggests that these two estimates are to be preferred over the others.  Note also that the scatter around the median is symmetric, whereas in the comparison with SDSS it is not.  

B11 argue that their sky estimates represent a substantial improvement over the standard SDSS catalog results and should form the basis of any analysis of nearby galaxies using the SDSS imaging data. Figure~\ref{blantonSky} shows that, in fact, this is not restricted to nearby galaxies:  E.g., for all galaxies with apparent sizes $\ge 7$ arcseconds, the SDSS sky is biased (left panel) (we quantify its effect on photometric parameters in Section~\ref{skydeV}). In the SDSS main galaxy sample these tend to be galaxies with large luminosities (right panel) which are typically in crowded fields.  The agreement between B11 and PyMorph in both panels suggests that, in contrast to the SDSS, PyMorph is unbiased for large luminous galaxies.

%B11 highlight the fact that the differences with respect to the SDSS pipeline values are driven by large nearby objects.  However, the agreement with PyMorph across the entire population shows that the differences with respect to SDSS are also pronounced at high luminosities, which tend to be galaxies in crowded fields that are not necessarily nearby.  

\subsection{Sky-related biases when fitting deVaucouleurs profiles to E+S0s}\label{skydeV}
Having determined that the PyMorph/B11 sky is to be preferred, we now consider how the choice of sky biases the inferred parameters.  We begin with a study of the only case in which a direct comparison (i.e. same model fit) with SDSS is possible:  fitting a deVaucouleurs profile to images of E+S0 galaxies. 
For morphological type, we use the Bayesian automated classifications of Huertas-Company et al. (2011):  each galaxy is assigned weights which represent the probabilities that it is Elliptical, S0, Sab or Scd. We restrict to E+S0s since a deVaucouleurs profile is known to not fit other morphological types well, and we do not wish to confound the question of sky-related biases with biases arising from fitting a bad model.
%In this case, there are really only two free parameters:  the half light radius, and the surface brightness at or within this radius.  

\begin{figure*}
 \centering
 \includegraphics[width = 0.45\hsize]{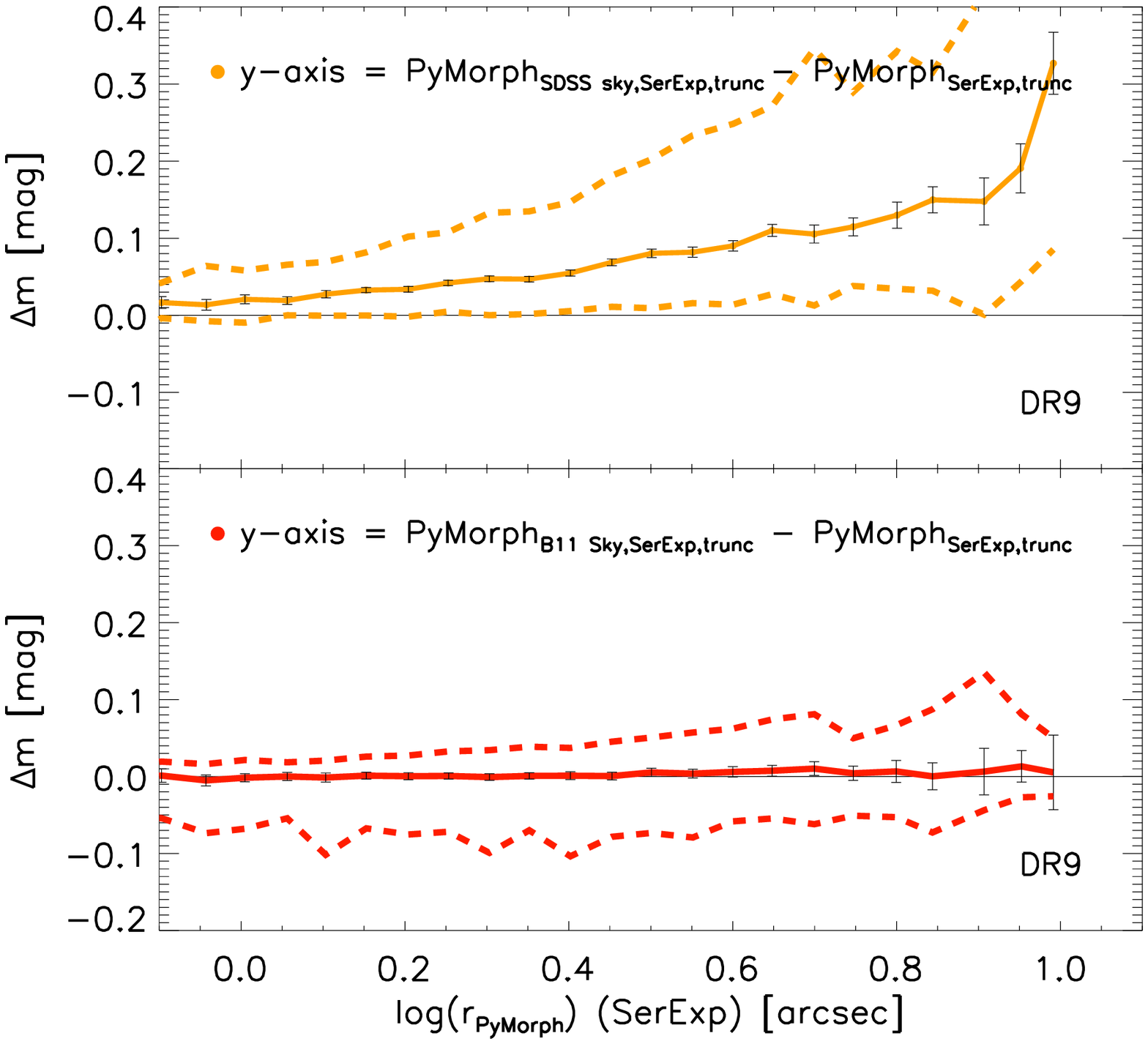}
 \includegraphics[width = 0.45\hsize]{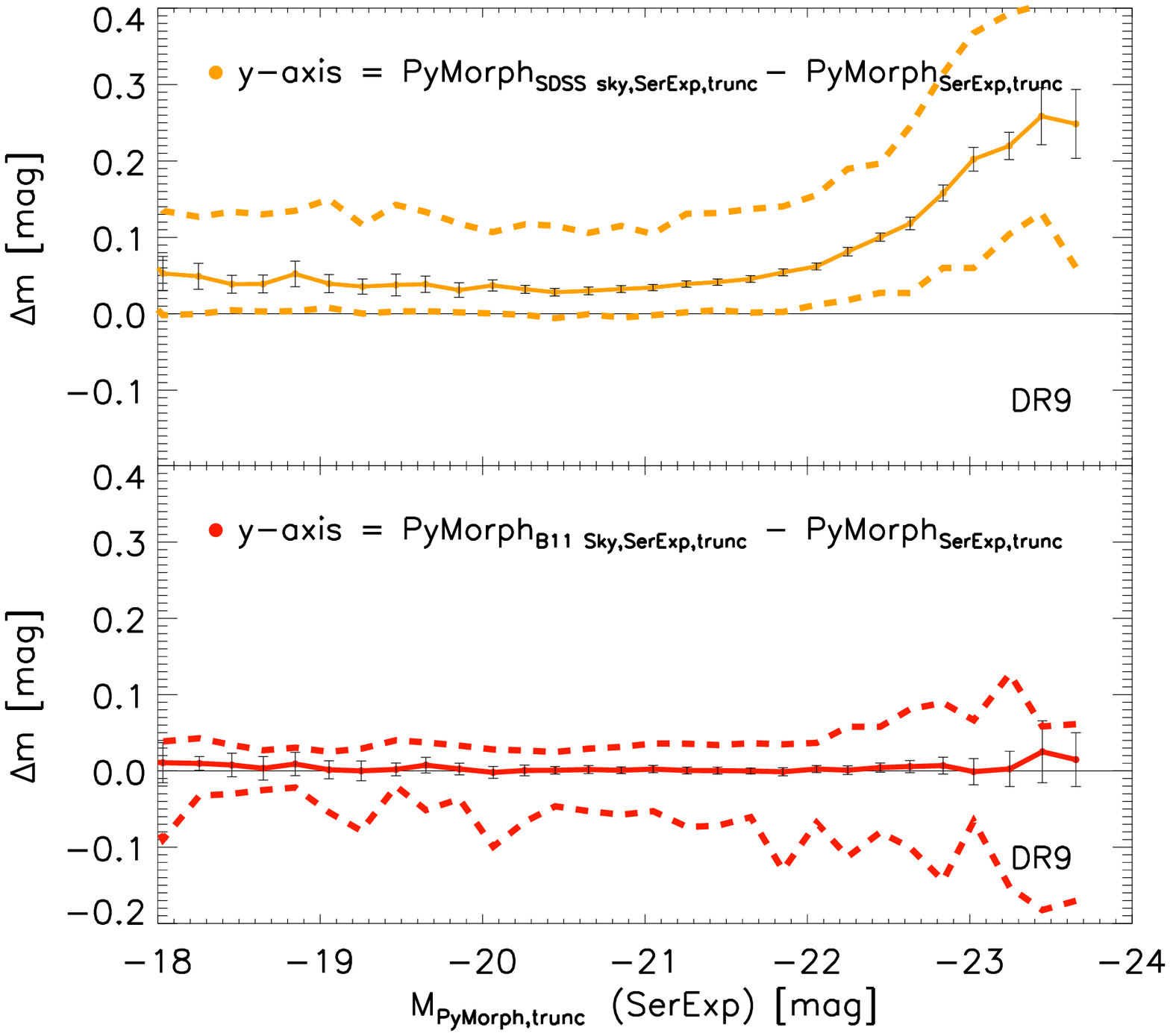}
 \includegraphics[width = 0.45\hsize]{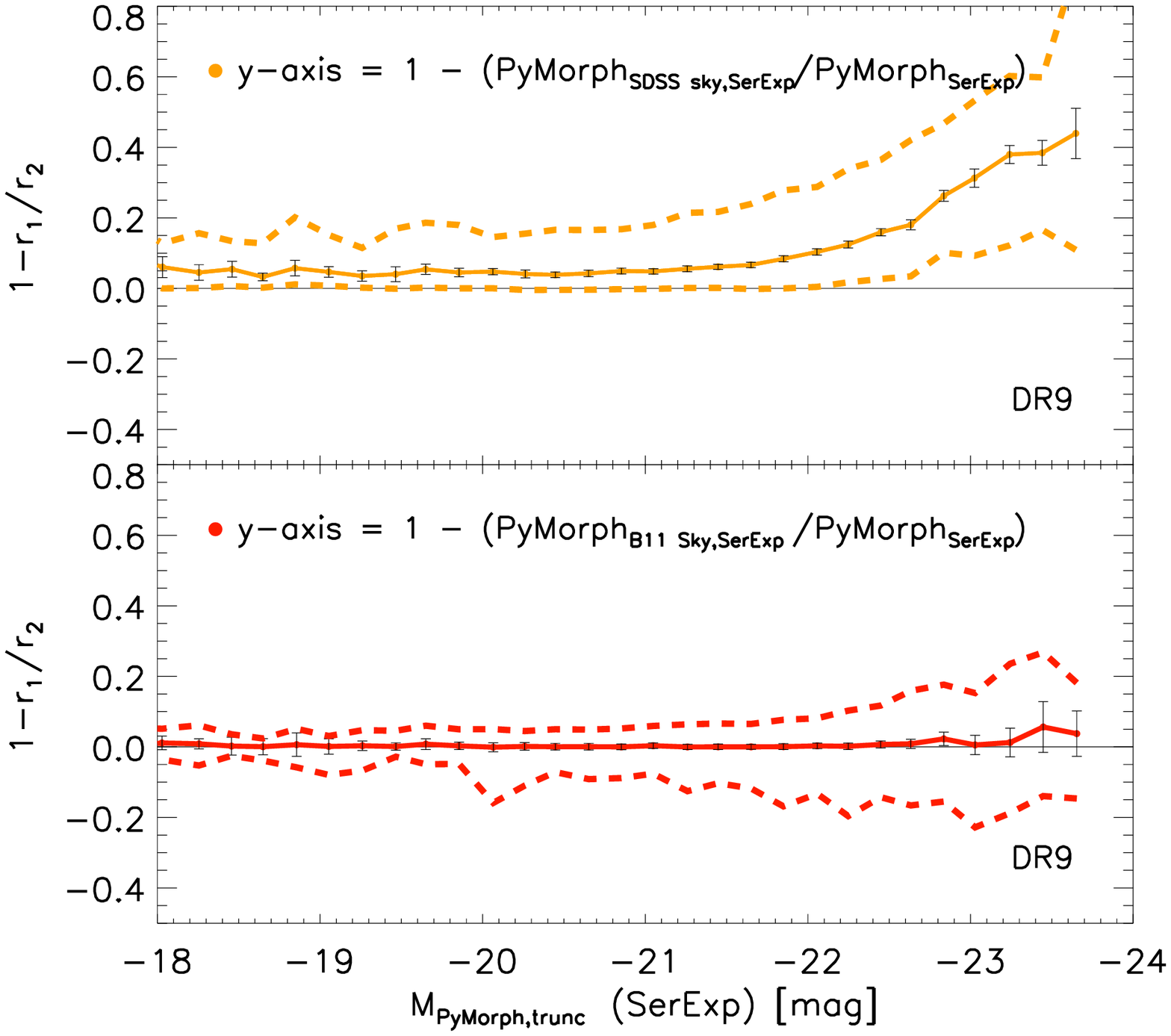}
 \includegraphics[width = 0.45\hsize]{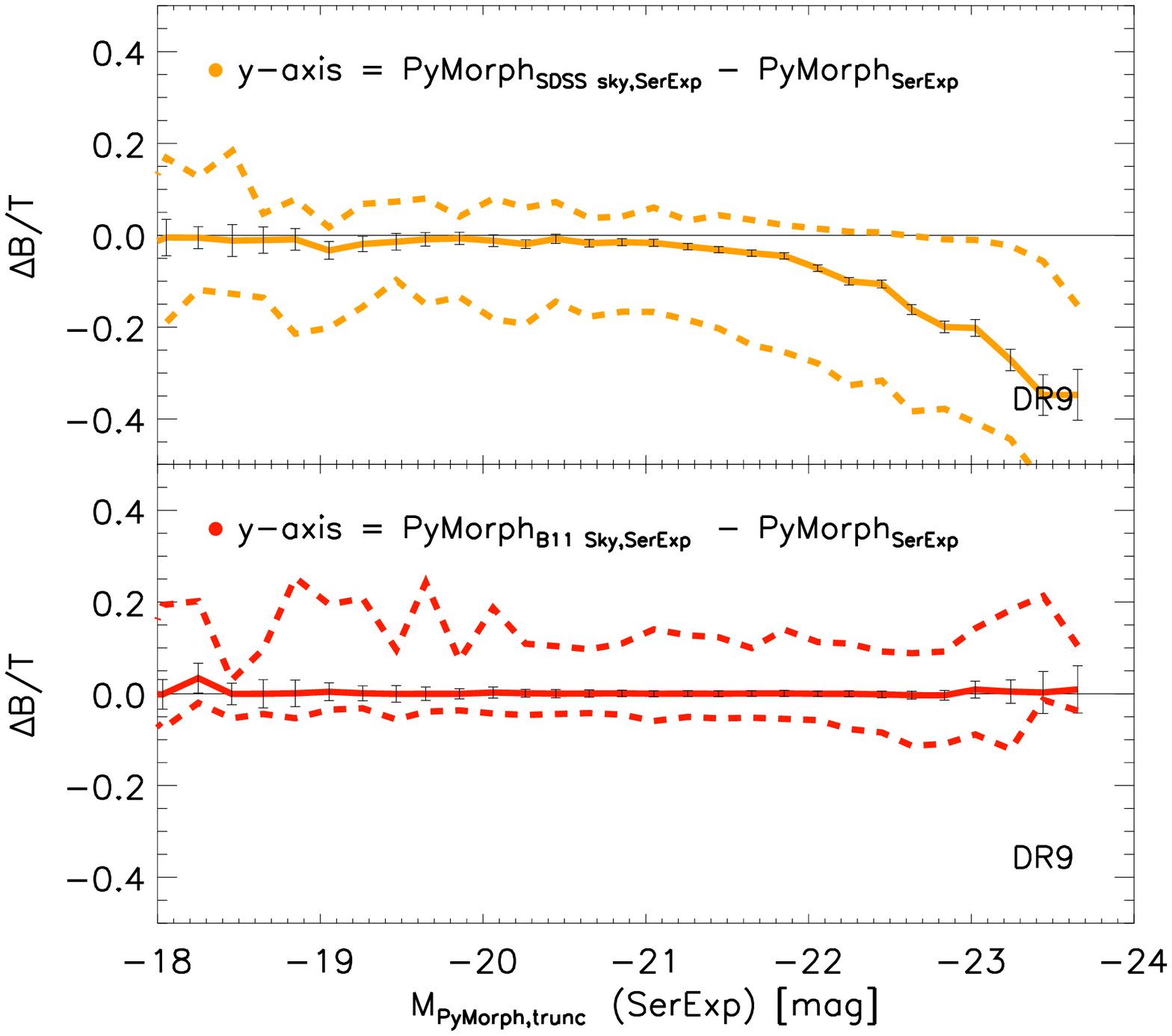}
 \caption{Sky-related biases associated with PyMorph DR9 SerExp fits to all galaxies (not just E+S0s).  Top halves of top two panels show that the brighter SDSS sky biases the estimated SerExp magnitude fainter, and this bias is most severe for the most luminous (left) and/or largest (right) galaxies.  Bottom halves show there is no bias if the Blanton et al. (2011; B11) sky is used.  Although the scatter is similar to that in the bottom left panel of Figure~\ref{deV}, degeneracies between the fitted SerExp parameters and the fitted sky contribute to increased scatter at high luminosities.  Bottom panels show that the SDSS sky leads to fainter magnitudes for large galaxies, and smaller sizes and B/T values for luminous galaxies, but that there are no such biases associated with the B11 sky values.}
 \label{serexp}
\end{figure*}

\begin{figure*}
 \centering
  \includegraphics[width = 0.45\hsize]{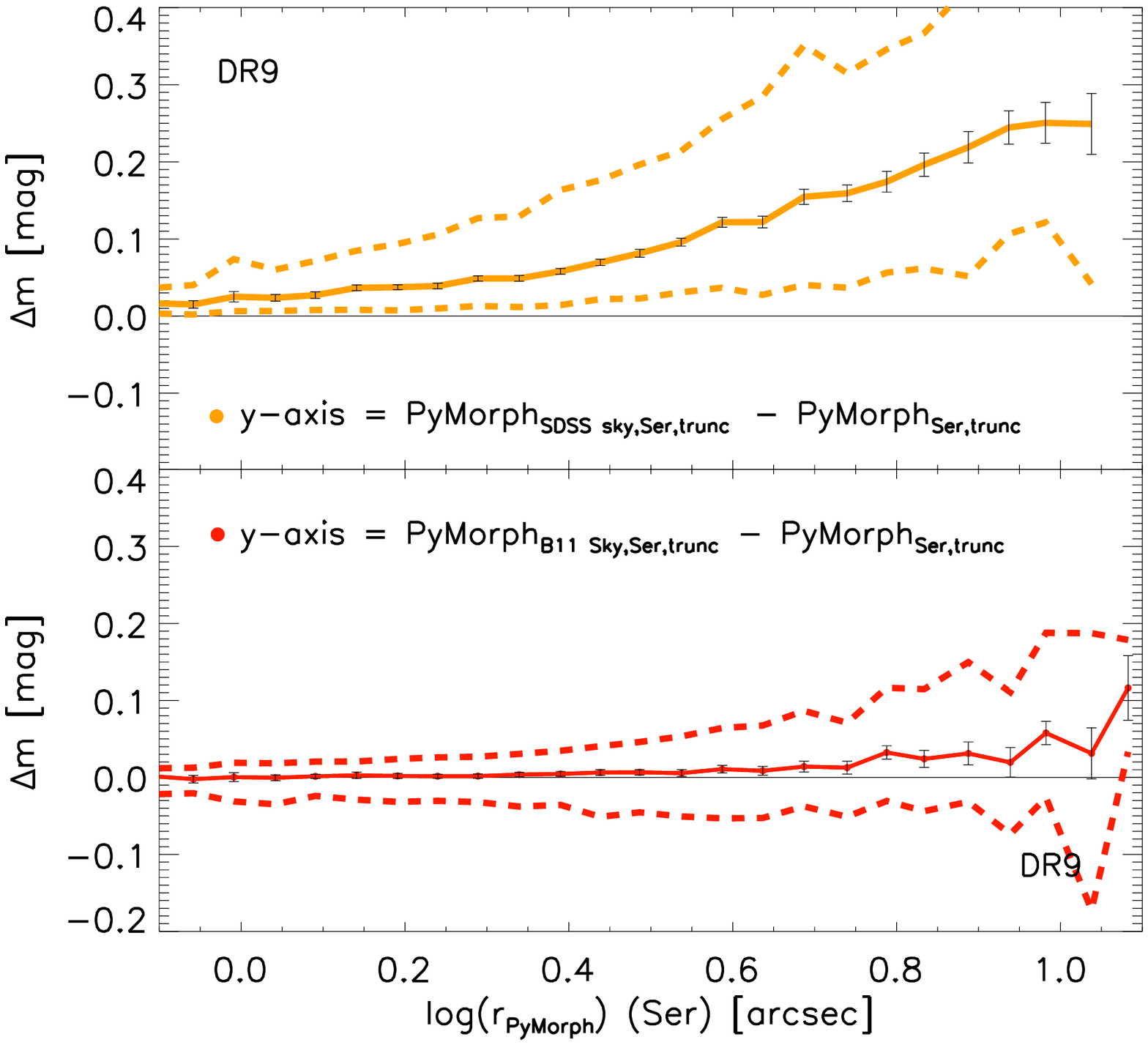}
  \includegraphics[width = 0.45\hsize]{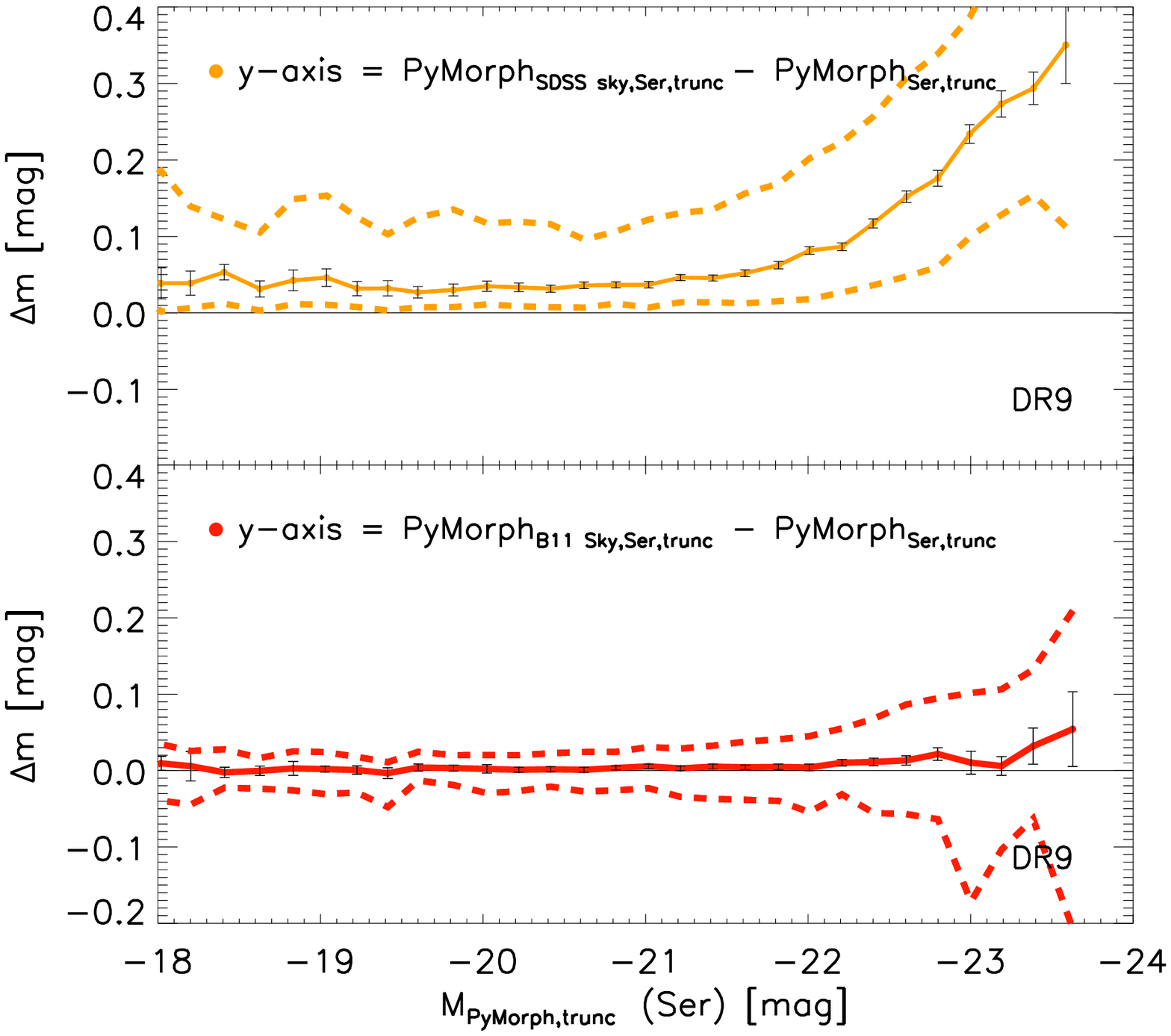}
  \includegraphics[width = 0.45\hsize]{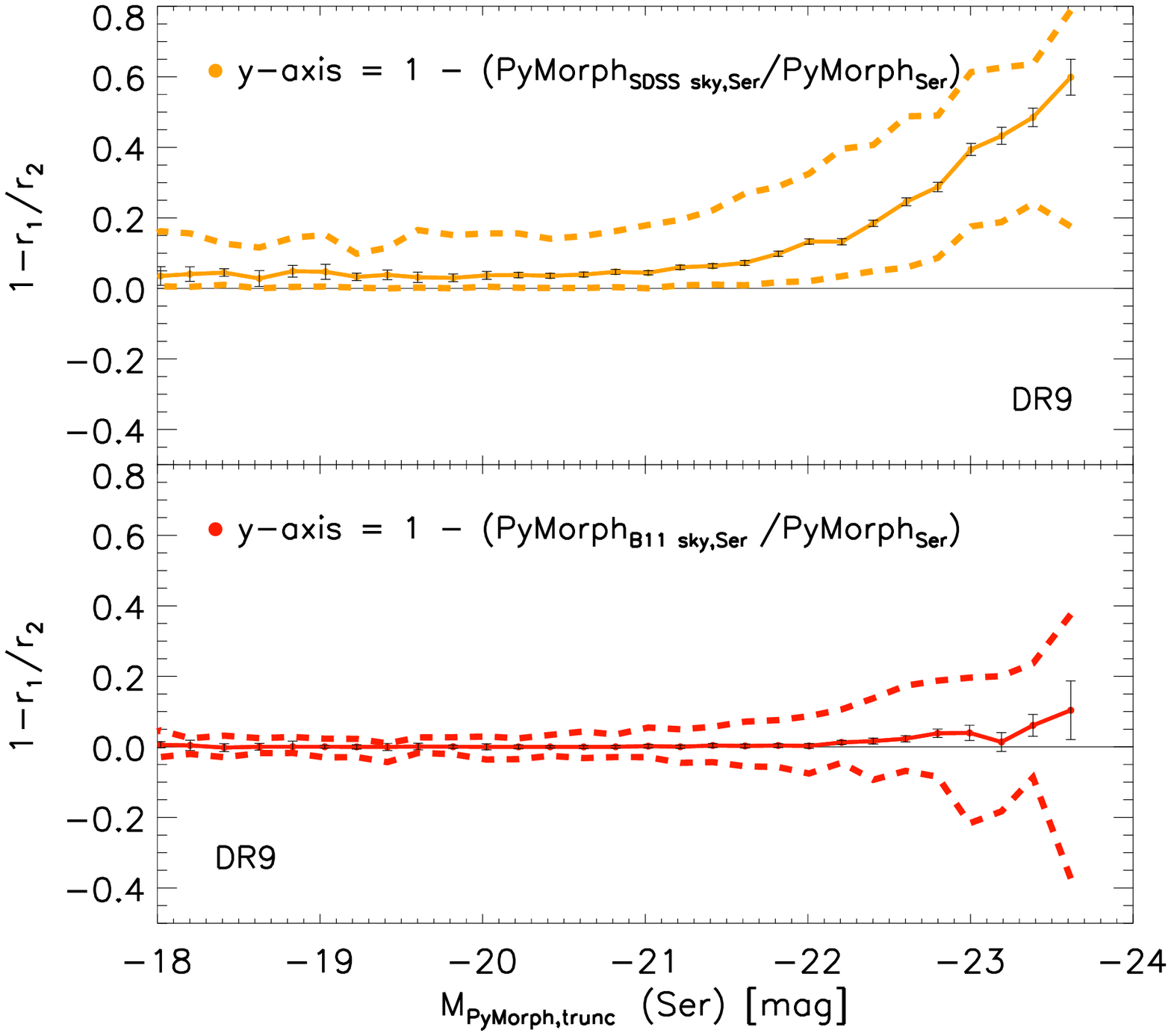}
  \includegraphics[width = 0.45\hsize]{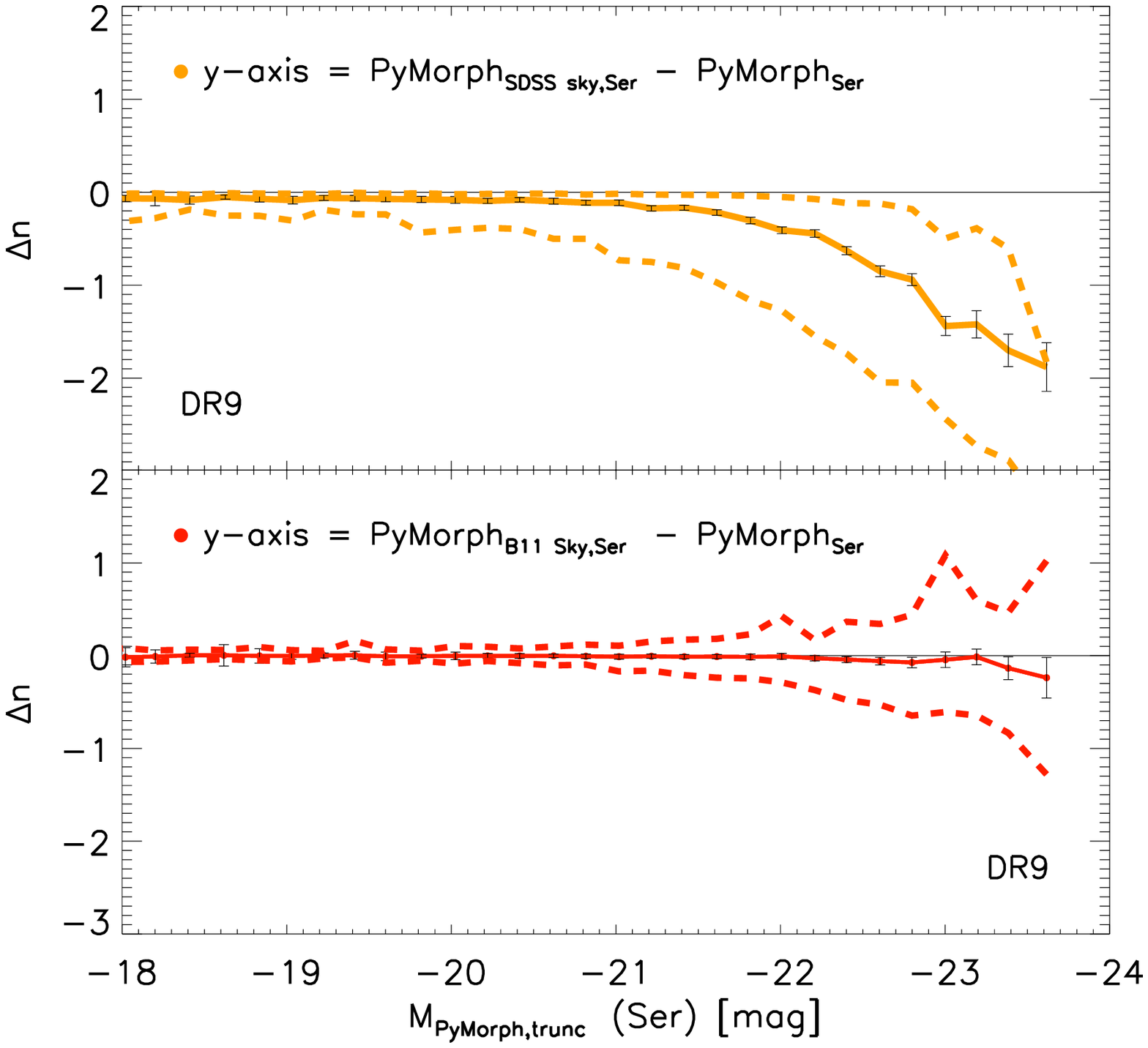}
 \caption{Same as previous figure, but now for Ser rather than SerExp fits to all galaxies, so bottom right panel shows the change in $n$ rather than B/T.  The SDSS sky biases the total light, half light radius, and $n$ towards lower values especially at large luminosities and angular sizes, but there are no such biases associated with the B11 sky values.}
 \label{ser}
\end{figure*}

The top panel of Figure~\ref{deV} shows the difference between the SDSS DR9 and PyMorph estimates of the total (truncated) magnitude. The black curve shows results for the DR9 E+S0 subset while the gray curve shows the results for the larger ($\sim 60\times$) PyMorph DR7 E+S0 sample. These curves show that SDSS is fainter, and this difference increases for the largest (left) and most luminous (right) galaxies.  This is a consequence of three effects:  (i) the SDSS sky is brighter, so galaxies with large angular radii tend to have their sizes reduced by a bigger factor, as a result of which less light is assigned to the galaxy; (ii) the total magnitude is computed by integrating the surface brightness profile, and our model of how the SDSS truncates this integral (equation~\ref{Ltrunc} and related discussion) may not be accurate; (iii) the SDSS and PyMorph fitting routines are systematically different.  

To remove the latter two effects, the yellow curve shows the difference between forcing PyMorph to use the SDSS sky values when fitting and the original PyMorph value.  Since both estimates are from PyMorph DR9, effects (ii) and (iii) have been removed, so the yellow curves differ from zero entirely because of the differences in sky values (the SDSS sky is brighter).  Moreover, the fact that these yellow curves agree with the previous black ones to better than 0.01~mags strongly suggests that we have modelled the SDSS truncation algorithm correctly:   PyMorph$_{\tt SDSSsky,deV,trunc}$ is a good proxy for SDSS$_{\tt deV}$.  (As an aside, this means that the good agreement at magnitudes fainter than $\sim -22$~mag in the top panels of Figures~\ref{pymorphSDSS7} and~\ref{pymorphSDSS9} is fortuitous, at least where the contribution from E+S0s is significant.)  

Figure~\ref{blantonSky} shows that while the SDSS sky is brighter than PyMorph, the B11 sky is in excellent agreement across the entire population.  The bottom panels of Figure~\ref{deV} show that if PyMorph is forced to use the B11 sky estimate rather than its own (in practice, this means PyMorph is made to fit the B11 sky-subtracted image provided on the SDSS website, while forcing its own additional sky estimate to be zero across the image), then the median difference in magnitude is negligible.  Notice that the scatter around the median is less than 0.03~mags; this level of agreement is remarkable.  
Comparison of the top and bottom panels shows that the sky can introduce biases of order 0.1~mags or more for the most luminous objects when fitting deVaucouleurs profiles.

\subsection{Sky-related biases in Ser and SerExp fits}
We now consider sky-related biases when fitting other models.  
Figure~\ref{serexp} shows results for PyMorph SerExp fits to all galaxies as the restriction to E+S0s is no longer necessary.  The yellow curves in the different panels show that the brighter SDSS sky biases the estimated SerExp magnitude fainter, and this bias is most severe for the largest (top left) and/or most luminous (top right) galaxies; it also biases the half-light radii and B/T values to smaller values (bottom left and right, respectively). For PyMorph SerExp fits the biases arising from the SDSS sky are significantly larger than when fitting deVaucouleurs profiles. However, there are no such biases associated with the B11 sky values (red curves).  On the other hand, although the scatter is similar to that in the bottom left panel of Figure~\ref{deV}, degeneracies between the fitted SerExp parameters and the fitted sky contribute to increased scatter at high luminosities. 

Figure~\ref{ser} shows a similar analysis of Sersic rather than SerExp fits.  In this case, there is only one component, so the bottom right panel shows the Sersic index $n$ rather than B/T.  Again, the SDSS sky biases the estimated Ser magnitude fainter, and this bias is most severe for the most luminous (top left) and/or largest (top right) galaxies; note that now the bias can be as large as 0.4~mags -- substantially larger than when fitting deVaucouleurs profiles.  The SDSS sky also biases the half-light radii and Sersic indices to smaller values (bottom left and right, respectively).  While there are no such biases associated with the B11 sky values, there are hints of a small bias at the largest angular sizes and luminosities. Since there are fewer free parameters compared to SerExp, and therefore fewer degeneracies, we would expect the scatter around the zero-median to be smaller. This is indeed the case for intermediate and low luminosity galaxies which usually have a Sersic index $n < 4$. A Sersic fit with a higher $n$ is more sensitive to differences in the background sky (Meert et al. 2013). Thus, the larger scatter observed at large sizes and/or luminosities is due to the fact that the most luminous galaxies usually have $n \ge 4$.

\begin{figure}
 \centering
 	\includegraphics[width = 0.9\hsize]{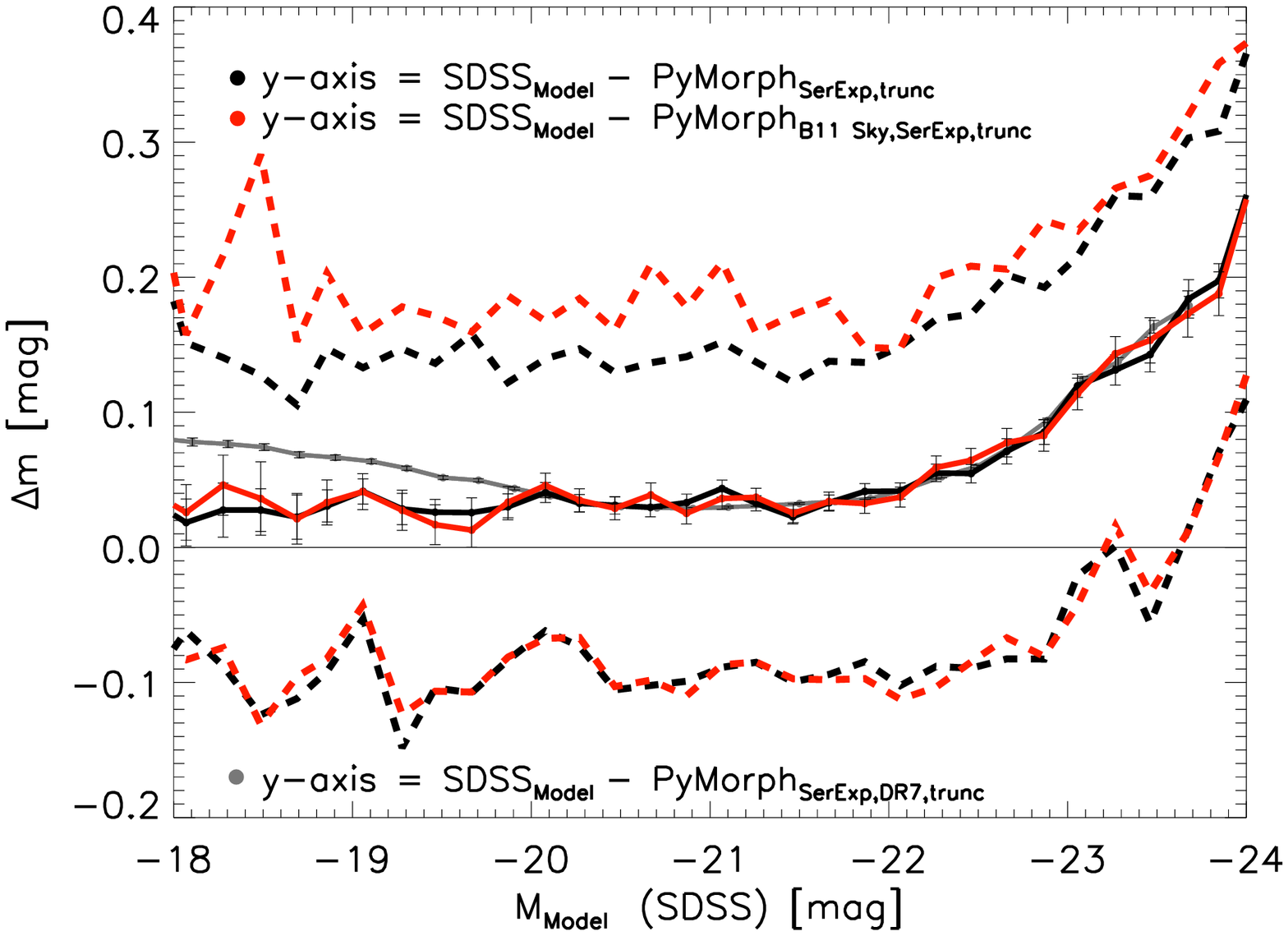}
 	\includegraphics[width = 0.9\hsize]{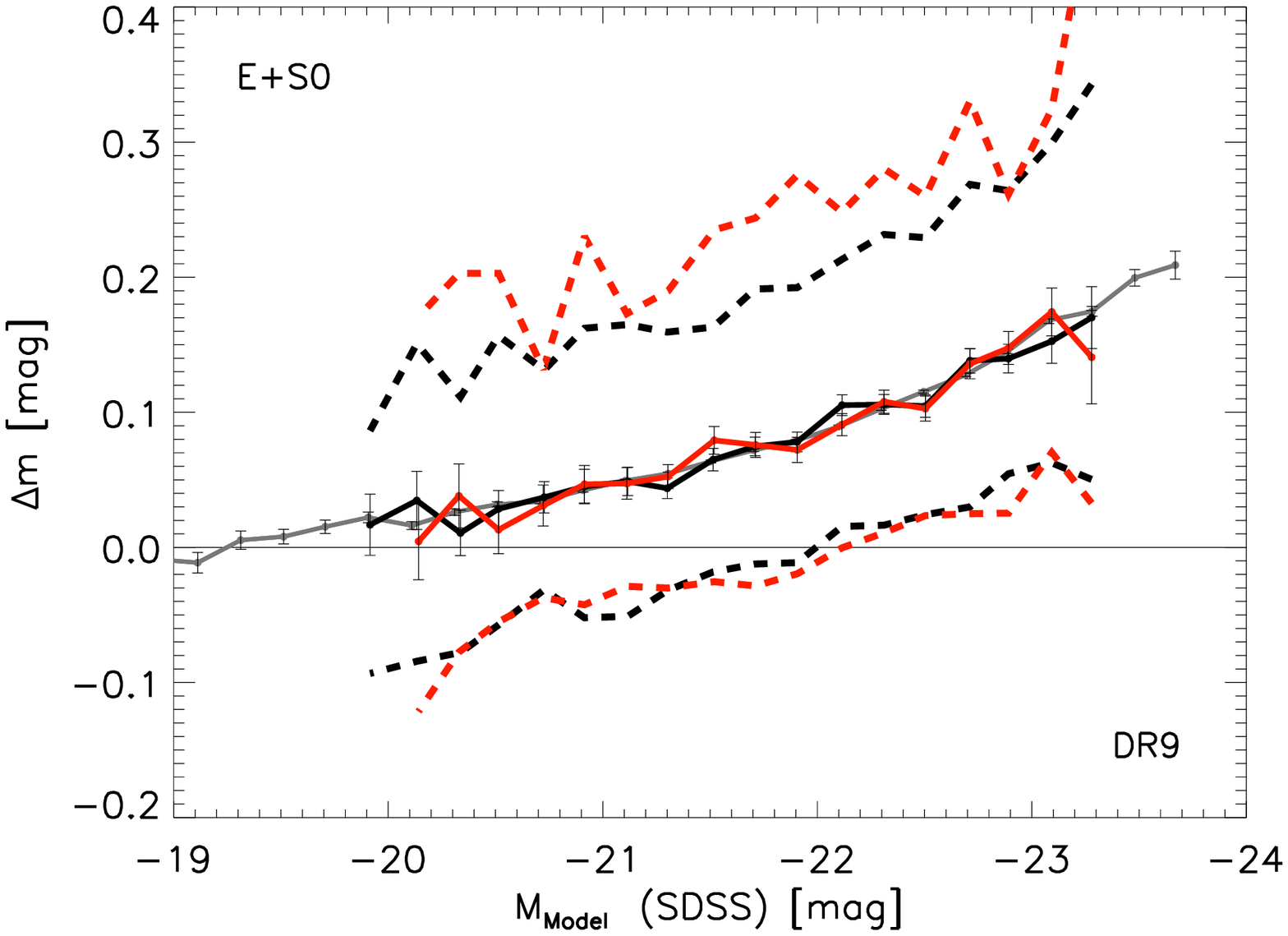}
\caption{Difference between SDSS DR9 {\tt Model} and PyMorph DR9 SerExp (truncated) magnitudes when PyMorph fits its own sky (black), and when the sky is fixed to that of Blanton et al. (2011; B11) (red) of the DR9 subset of $10^4$ galaxies.  Top panel shows results for all galaxies; bottom panel shows the subset of galaxies classified as E+S0s.  The gray curves in the two panels show the comparison between SDSS DR9 {\tt Model} and PyMorph DR7 SerExp magnitudes of the much larger ($\sim 60\times$) sample.}
 \label{dsouza}
\end{figure}

The results of this subsection have an interesting connection to recent work.  D'Souza et al. (2015) state that image stacking is essential for recovering unbiased estimates of the total light.  Their stacks were of DR9 sky subtracted images, meaning that they assumed the B11 sky estimate was correct.  The results in the bottom halves of each panel in Figures~\ref{deV}--\ref{ser} were based on analyses of individual images. Since no stacking was performed when fitting, the lack of bias between the full PyMorph values and those when the sky is fixed to that of B11 shows that stacking is not a prerequisite for obtaining unbiased results.

In this context, it is interesting to compare the difference between PyMorph SerExp and SDSS {\tt Model} magnitudes.  Bernardi et al. (2017a) have already shown that the median difference is the same as what D'Souza et al. find from their stacking analyses (see their Figure~2).  But they left open the question of the scatter around the median.  Since their work used SerExp magnitudes in which PyMorph also fit for the sky, it is possible that some of the scatter is reduced when PyMorph is forced to use the B11 sky.  The top panel of Figure~\ref{dsouza} shows that this is not the case:  the differences between PyMorph SerExp and SDSS {\tt Model} magnitudes when PyMorph fits its own sky and when the sky is fixed to that of B11 are very similar, not just in the median but also the scatter around it.  (Because we are using truncated magnitudes, comparing PyMorph to SDSS quantities, the offset from zero is due to differences in sky and fitted-model only.)  This strongly suggests that the scatter reflects true differences between SerExp and Model (i.e. deV) models; it is not dominated by degeneracies arising from fitting the sky simultaneously.  To remove trends which arise from morphology, the bottom panel shows a similar analysis for the subset of galaxies classified as E+S0s.  While the trends differ especially at low luminosities -- where non-E+S0s begin to dominate in the top panel -- it is still true that changing from PyMorph to B11 sky values makes little difference.

\section{Dependence on fitted model}\label{Pmodels}
Having shown the large biases associated with the SDSS, we now turn exclusively to PyMorph values.  Recall that the PyMorph sky values are essentially the same for all fitted models (Figure~\ref{sameSky}), so that comparison of different PyMorph fits show how the luminosity and size depend on the functional form assumed for the surface brightness profile.  Also, when comparing results from deVaucouleurs profiles we show results for E+S0s only to avoid the issue of biases which arise from using a functional form which is known to provide a poor fit.  

Figure~\ref{dev2serexp7} shows that SerExp fits to E+S0s return more light than deV fits especially at large luminosities (up to $\sim 0.2$~mag); when shown as a function of SerExp luminosity, the difference is largest for the most luminous galaxies.  (This analysis was done using the DR7 E+S0s, for which PyMorph reductions are available, since it is much larger ($\sim 60\times$) than the subset of DR9 galaxies on which PyMorph was rerun. The result from the DR9 subset is noisier, but otherwise very similar, so we have not included a separate figure showing it.)

The difference due to fitting different models is similar in amplitude to that in Figure~\ref{deV}, which was due to differences in the estimated sky.  However, the dependence on choice of regression is more dramatic here than in Figures~\ref{curved1} and~\ref{curved2} because there the effects of the sky somewhat compensated for the difference in profiles.  By using only PyMorph quantities here, the sky effects have been removed.  

Figure~\ref{serexp2ser7} shows a similar comparison, but now between SerExp and Ser fits to DR7 E+S0s. Clearly, SerExp is about 0.1~mags fainter and 10\% smaller across the E+S0 population.  Finally, Figure~\ref{serexp2serAll7} compares SerExp and Ser fits to the full DR7 population.  At high luminosities, this figure is very similar to the previous one, because most high luminosity galaxies are E+S0s.  However, there are small differences at low luminosities.  These indicate that Sersic luminosities and sizes of non-E+S0s must be fainter and smaller than the corresponding SerExp values.  

The differences between the cyan and magenta curves in Figures~\ref{dev2serexp7} and~\ref{serexp2ser7} at the high luminosity end strongly suggest that the most luminous galaxies have different surface brightness profiles from the bulk of the population.  This can be understood as follows.  Suppose we have two populations, both of which span the same range of {\tt deV}.  Assume that, for one, {\tt Ser}$=${\tt deV}, but that {\tt Ser}$=${\tt deV}$-m$ for the other (i.e. {\tt Ser}\ is $m$~mags brighter).  Let $f$ denote the fraction of objects in this second population.  Then, a plot of $\Delta M \equiv {\tt deV}-{\tt Ser}$ when shown as a function of {\tt deV}\ will look like two horizontal lines, one lying $m$~mags above the other, like this:  $=$.  The average of $\Delta M$ when shown as a function of {\tt deV}\ will equal $fm$.   However, when shown as a function of {\tt Ser}, the second population will be displaced brightwards along the x-axis:  \underline{~~}$-$.  As a result, where the two populations overlap, the mean $\Delta M$ will still be $fm$, but at the brightest {\tt Ser}\ the mean will be $m$.  If the second population only spans a limited range of {\tt deV}, $-=-$, then the average as a function of {\tt deV}\ will show curvature, and may not even be monotonic, whereas the other may still be monotonic: $--=$.  Alternatively, suppose that when plotted as a function of {\tt Ser}, $\Delta M$ is made of two populations:  {\underline{~~} $|$.  Then, when plotted as a function {\tt deV}, this will look like \underline{~~}\textbackslash.  Again, the plot versus {\tt Ser}\ will be monotonic, whereas that versus {\tt deV}\ will not.  In practice, this mix of populations means that if one wishes to use the mean of $\Delta M$ as a measure of the difference between {\tt deV}\ and {\tt Ser}, then one {\em must} specify which variable was being held fixed (we made a similar point in the context of Figures~\ref{curved1} and~\ref{curved2}).
  
%Figures~\ref{serexp2ser9} and~\ref{serexp2ser7} show a similar comparison, but now between SerExp and Ser fits to E+S0s.  Clearly, SerExp is about 0.1~mags fainter and 10\% smaller across the E+S0 population.  Finally, Figures~\ref{serexp2serAll9} and~\ref{serexp2serAll7} compare SerExp and Ser fits to the full population.  Comparison with the previous figures suggests that, for non-E+S0s, Sersic luminosities and sizes must be about 0.1~mags fainter and 10\% smaller at low luminosities.

\section{Conclusions}\label{final}
In both SDSS DR7 and DR9, PyMorph returns brighter estimates of the total light of a galaxy than either SDSS {\tt Model} or {\tt cModel} magnitudes (Figures~\ref{pymorphSDSS7} and~\ref{pymorphSDSS9}).  While the SDSS values have changed slightly between DR7 and DR9, the PyMorph fits to the DR7 release provided by Meert et al. (2015, 2016) remain accurate for DR9 as well (Figures~\ref{dr7to9ser} and~\ref{dr7to9serexp}).  Some of the difference with respect to the SDSS arise from the fact that the SDSS value for the total brightness comes from truncating the integral over the surface brightness profile (Figures~\ref{ftruncate}, \ref{curved1}, and \ref{curved2}).  We believe we understand the truncation algorithm (Figure~\ref{deV} and related discussion), and so in all our subsequent comparisons with the SDSS, we have truncated the PyMorph values using a similar algorithm (equation~\ref{Ltrunc}) so that truncation plays no further role in the PyMorph-SDSS differences.  

The sky estimated by PyMorph is almost completely independent of the model used to fit the galaxy (Figure~\ref{sameSky}).  The PyMorph sky estimates are fainter than those of the SDSS DR7 or DR9 pipelines (Figure~\ref{SDSS2sky}), but are in excellent agreement with the estimates of B11 (Figure~\ref{blantonSky}).  The difference in sky accounts for about half of the discrepancy shown in Figures~\ref{pymorphSDSS7} and~\ref{pymorphSDSS9}.  In addition, there is an overall offset of about 0.07~mags which comes from the fact that the SDSS value for the total brightness comes from truncating the integral over the surface brightness profile (Figure~\ref{curved2}).  The remainder arises from fitting different models.

\begin{figure}
 \centering
 	\includegraphics[width = 0.9\hsize]{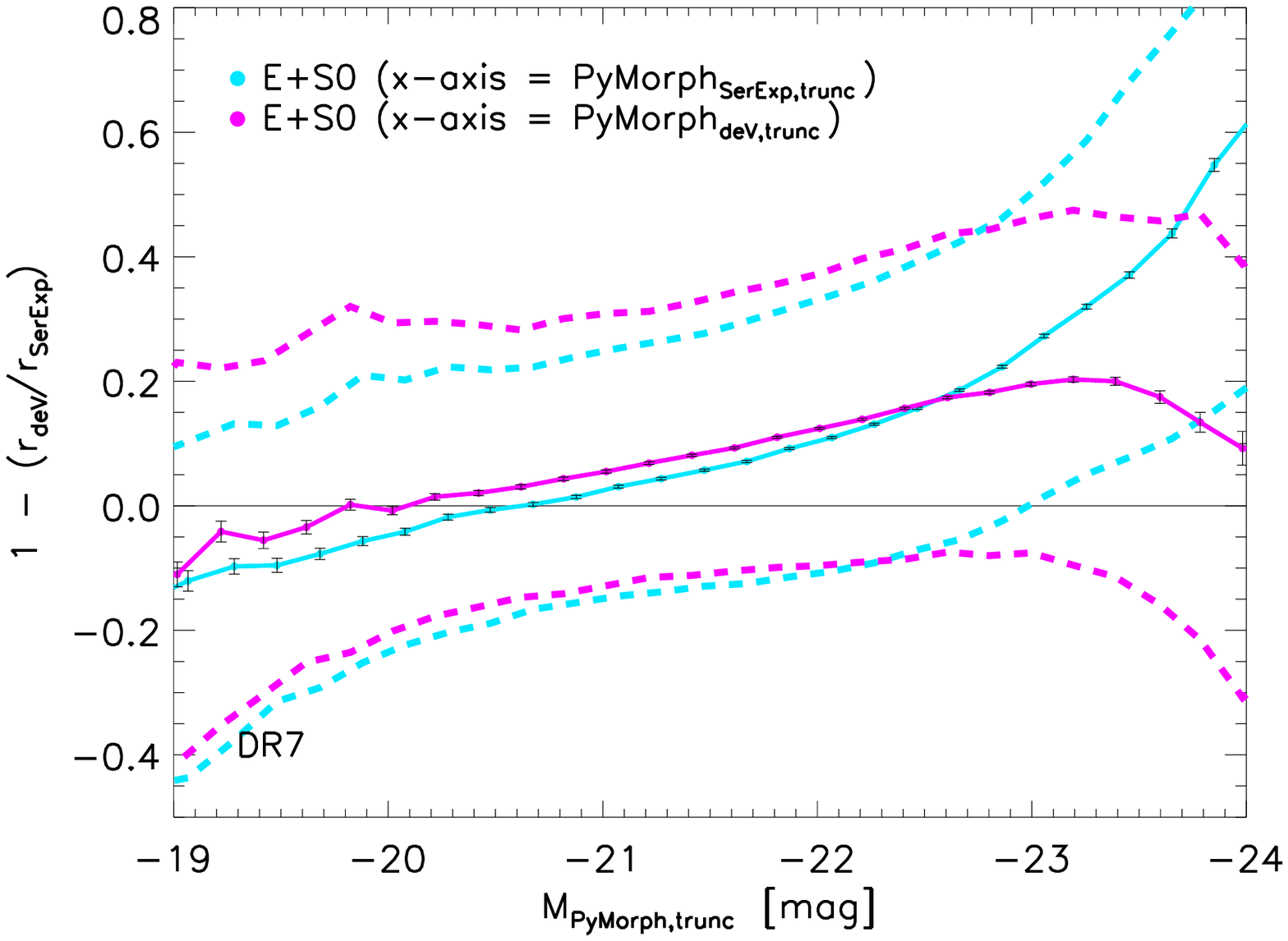}
 	\includegraphics[width = 0.9\hsize]{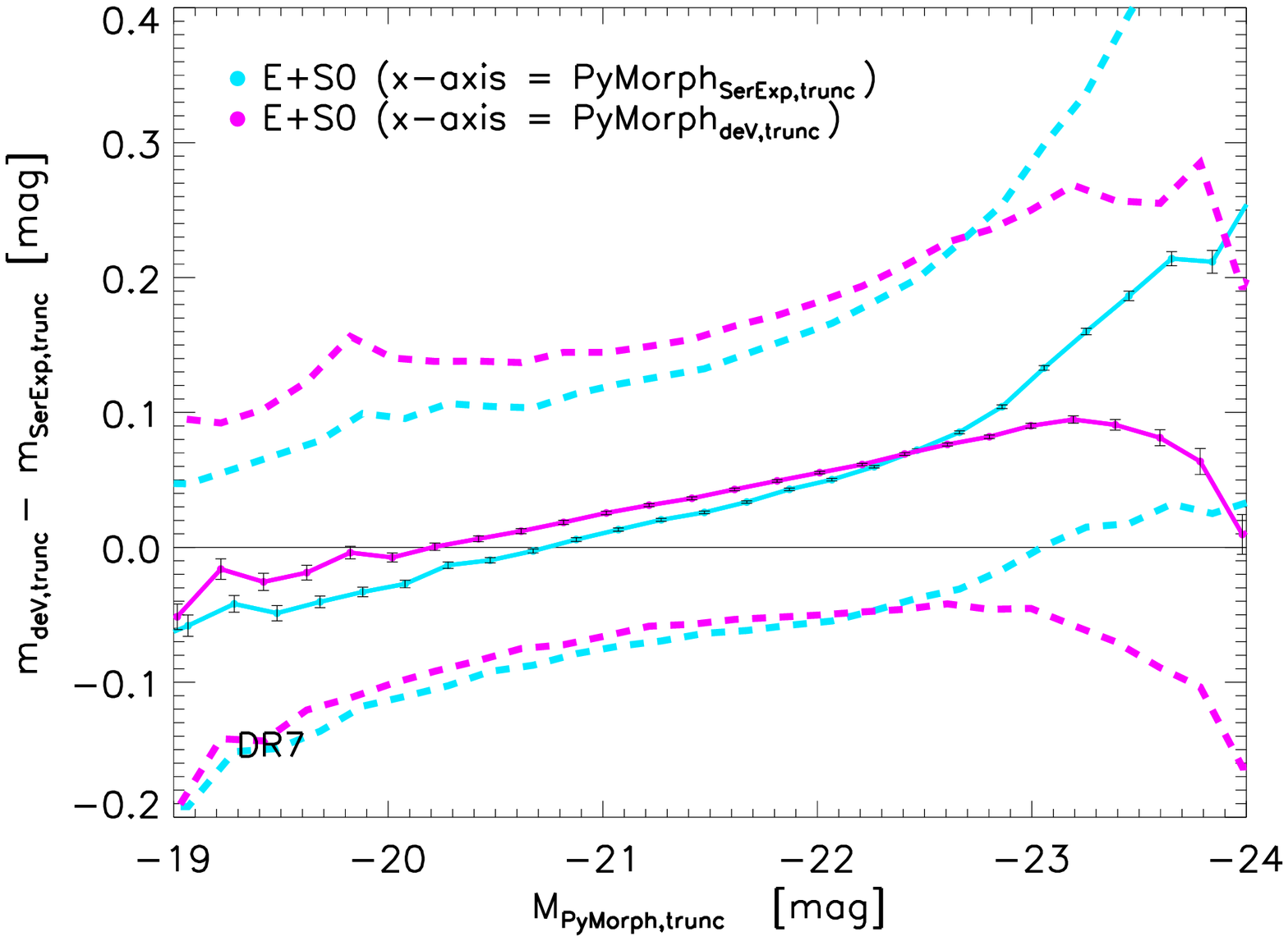}
\caption{Comparison of the half-light radii (top) and the total light (bottom) returned by PyMorph when fitting deVaucouleur and SerExp models E+S0s.  SerExp fits tend to be brighter and larger. At the brightest luminosities the difference between SerExp and deV is large enough that plotting versus one or the other makes a significant difference. When shown as a function of SerExp luminosity, the difference is largest for the most luminous galaxies.  Figure~\ref{sameSky} shows that the sky is essentially the same for these fits, so the differences are almost entirely due to the increased freedom which the SerExp model has compared to deV.}
 \label{dev2serexp7}
\end{figure}

Use of the SDSS sky biases luminosities and half-light radii to lower values; in the main SDSS galaxy sample these biases are significant (a few tenths of a magnitude) at large luminosities:  they matter not just for nearby galaxies.  The biases become even worse when allowing the model more freedom to fit the surface brightness profile (Figures~\ref{deV}--\ref{ser}).  

When PyMorph sky values are used, the SerExp fits to E+S0s return more light than deV fits especially at large luminosities (up to $\sim 0.2$~mag), but less light than Ser fits (Figure~\ref{serexp2ser7}).  For non-E+S0s, which are dominant towards lower luminosities, Sersic luminosities and sizes are slightly fainter and smaller than SerExp (Figure~\ref{serexp2serAll7}).  

Our findings show that, especially at large luminosities, SDSS pipeline values should not be used:  PyMorph estimates are much more reliable.  Of these, Meert et al. (2013) and Bernardi et al. (2014) have already shown that the SerExp values are to be preferred.  The PyMorph SerExp values are also consistent with results obtained via the stacking analysis of D'Souza et al. (2015) (Figure\ref{dsouza}; see also Figure~2 in Bernardi et al. 2017a). This is reassuring because the two analyses are very different.  However, this does raise the question of why SerExp is better than SDSS pipeline photometry.  E.g., since the largest discrepancies occur at high luminosities, and the most luminous galaxies are preferentially found in clusters, is it possible that the SerExp fits are different because the second component is actually fitting intercluster light?  Bernardi et al. (2017b) show that for the vast majority of massive galaxies this is almost certainly not the main reason for the difference.

The assembly history of a galaxy is expected to leave an imprint on its surface brightness profile. Indeed, we find significant evidence that the surface brightness profiles of the most luminous galaxies suggest that they are a distinct population (Figures~\ref{curved1}, \ref{curved2}, \ref{dev2serexp7} and~\ref{serexp2ser7}).  Therefore, we hope our results will inform studies of the assembly histories of the most massive galaxies.

\begin{figure}
 	\includegraphics[width = 0.9\hsize]{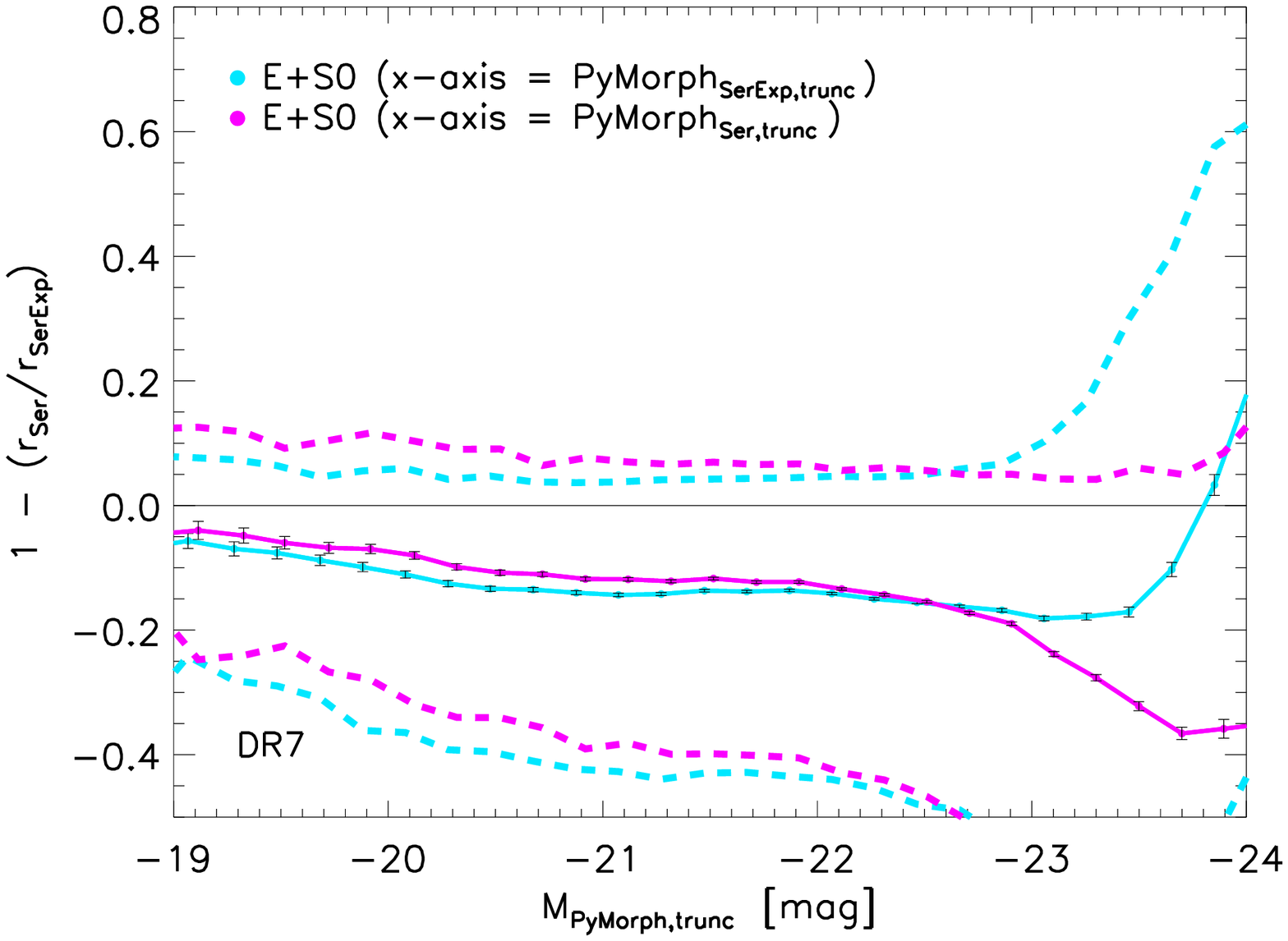}
 	\includegraphics[width = 0.9\hsize]{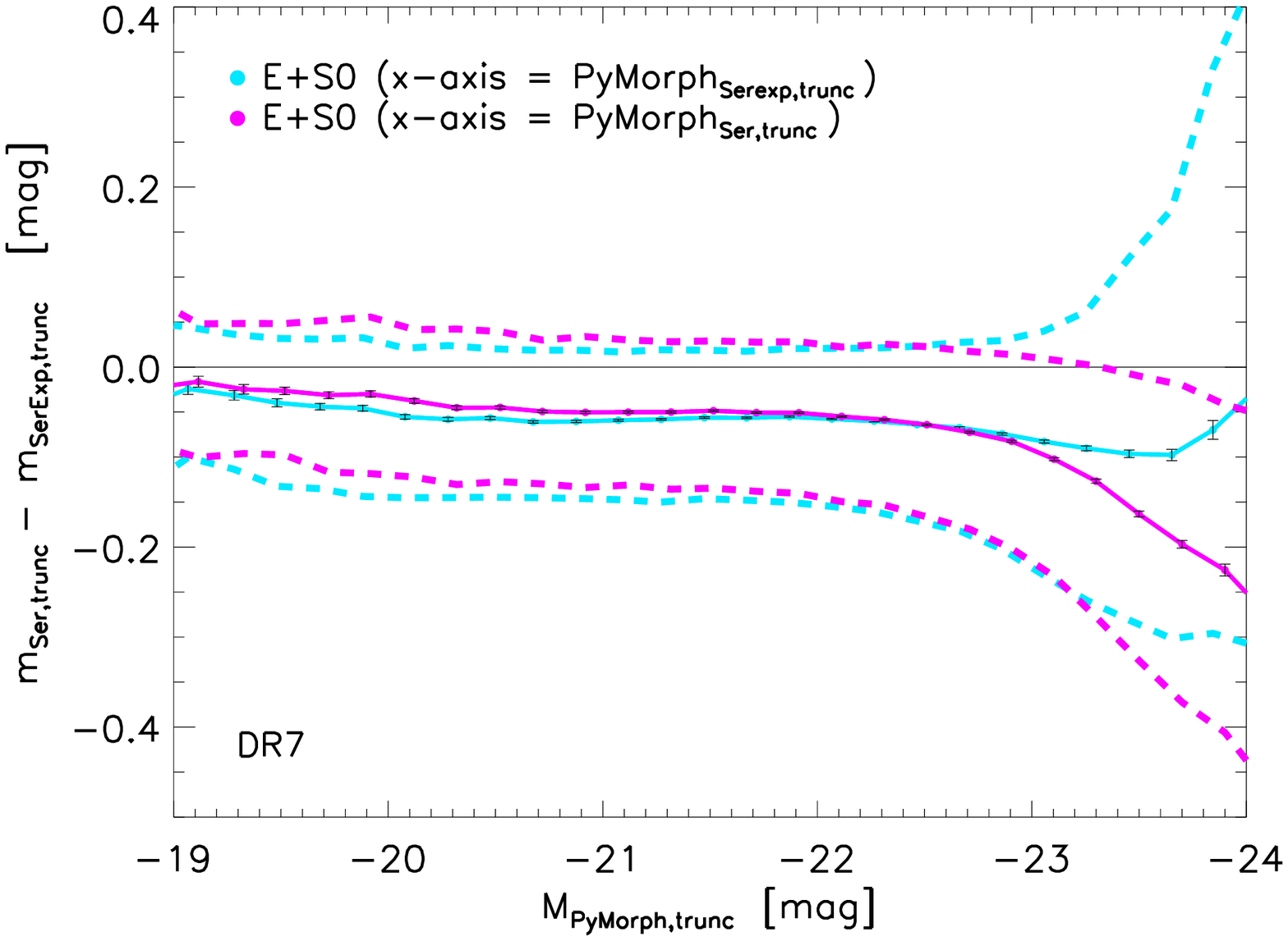}
\caption{Same as previous figure, but now comparing PyMorph SerExp and Ser fits for E+S0s. SerExp tends to be slightly smaller and brighter; at the brightest luminosities the difference is larger so plotting versus Ser or SerExp luminosity matters.}
 \label{serexp2ser7}
\end{figure}

\begin{figure}
 \centering
 	 \includegraphics[width = 0.9\hsize]{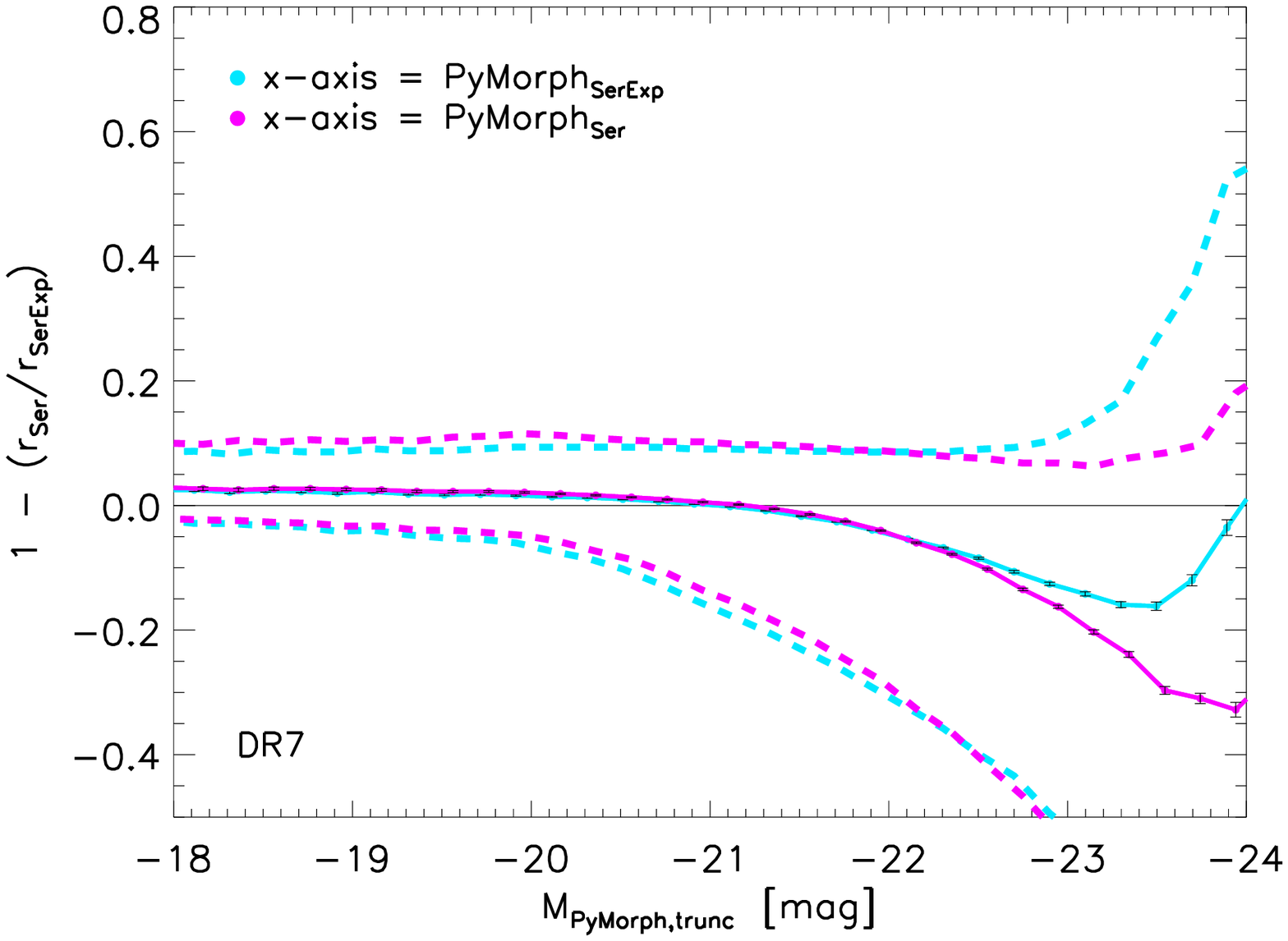}
 	 \includegraphics[width = 0.9\hsize]{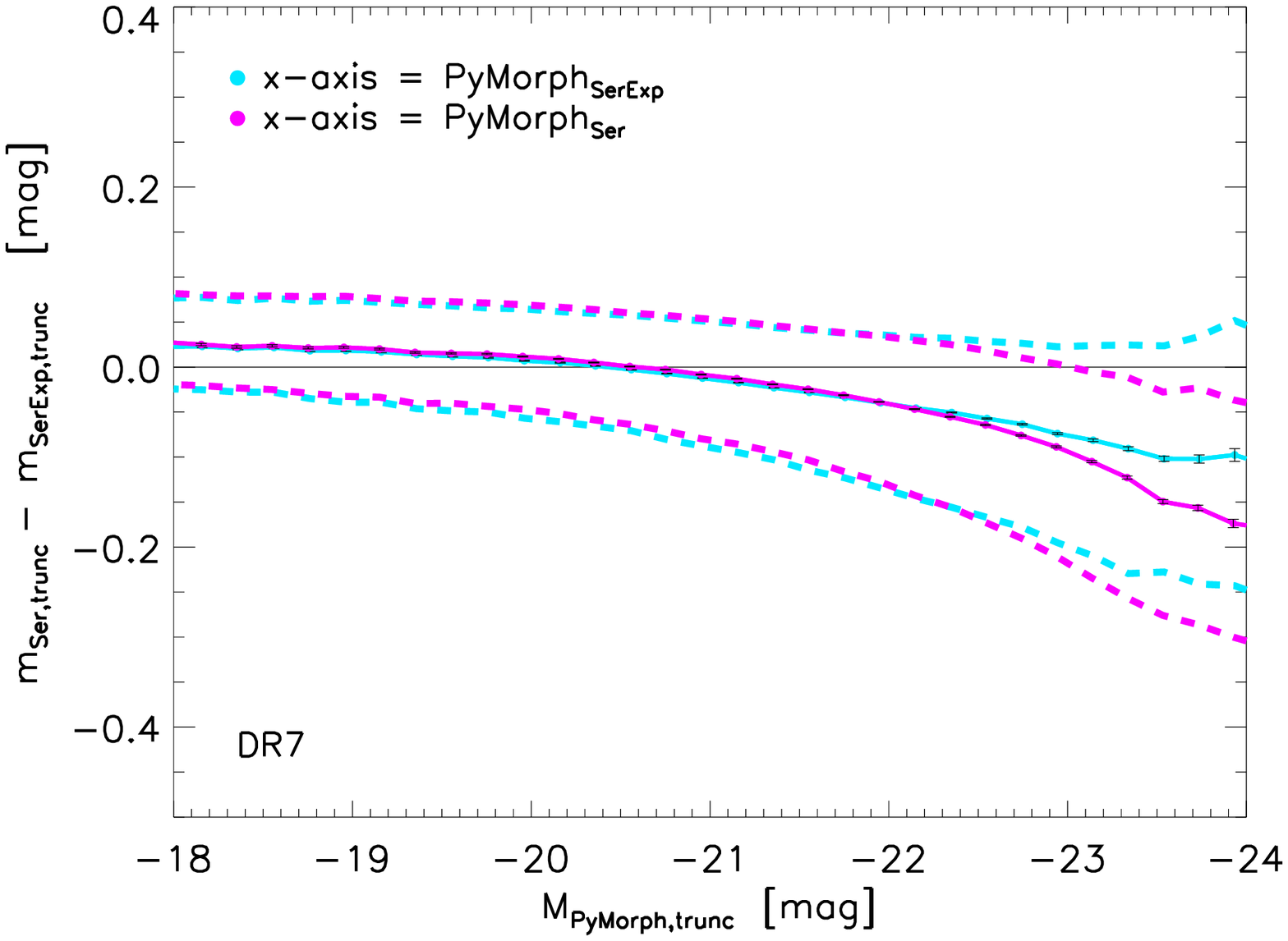}
\caption{Same as previous figure, but now for all galaxies.  Since most high luminosity galaxies are E+S0s, comparison with the previous figure shows differences only at low luminosities:  for non-E+S0s, Sersic luminosities and sizes tend to be fainter and smaller than their corresponding SerExp values.}
 \label{serexp2serAll7}
\end{figure}

\subsection*{Acknowledgements}
We thank R.K. Sheth and V. Vikram for helpful discussions, and the referee for a helpful and competent report.

\end{document}